\documentclass{article}%
\usepackage{amssymb}
\usepackage{amsmath}
\usepackage{makeidx}
\usepackage{amsfonts}
\usepackage{graphicx}
\usepackage{epsfig}%
\include{psfig.tex}
\usepackage{cite}%
\begin{document}

\author{Hartmut Wachter\thanks{E-Mail: Hartmut.Wachter@gmx.de}\\An der Schafscheuer 56\\D-91781 Wei\ss enburg, Federal Republic of Germany}
\title{Nonrelativistic one-particle problem on $q$-de\-formed Euclidean space}
\maketitle
\date{}

\begin{abstract}
We consider time-depen\-dent Schr\"{o}\-dinger equations for a free
nonrelativistic particle on the three-di\-men\-sion\-al $q$-de\-formed Euclidean space. We
determine plane wave solutions to these Schr\"{o}\-dinger equations and show
that they form a complete orthonormal system. We derive $q$-de\-formed expressions
for propagators of a nonrelativistic particle. Considerations about expectation values for position or
momentum of a nonrelativistic particle conclude our studies.

\end{abstract}

\section{Introduction}

Describing space-time as a continuum is a very successful concept in physics.
Various considerations, however, suggest that the Planck length limits the
accuracy of spatial measurements \cite{Mead:1966zz,Garay:1995}. This
uncertainty in space-time suggests that describing space-time as a continuum
is no longer appropriate for small space-time distances.\footnote{In his
inaugural lecture, Bernhard Riemann had already considered the necessity to
modify the geometry of space if spacings get smaller and smaller.}

In quantum mechanics, Heisenberg's commutation relations for position
coordinates and momentum coordinates imply that the state of a particle in
phase space is not accurately determined. We can assume analogously that the
uncertainty in position measurements is due to a noncommutativity of position
operators:%
\begin{equation}
\lbrack\hspace{0.01in}\hat{x}^{i},\hat{x}^{j}]=\theta^{ij}(\hat{x}).
\label{VerRelKooAlg}%
\end{equation}

H.~Snyder was one of the first who tried to construct a quantized space-time
with non-com\-mut\-ing position coordinates \cite{Snyder:1947a}. In recent times
the focus has been on noncommutative space-time algebras with $\theta^{ij}%
\in\mathbb{C}$
\cite{Doplicher:1994zv,Chu:1998qz,Schomerus:1999ug,Grimstrup_2002}.
Furthermore, space-time algebras with the commutator in
Eq.~(\ref{VerRelKooAlg}) being a linear function of space-time coordinates,
i.~e. $\theta^{ij}(\hat{x})=\Theta_{k}^{ij}\hspace{0.01in}\hat{x}^{k}$, are
also of particular interest \cite{Lukierski:1991pn,Majid:1994cy}. We consider,
however, the $q$-de\-formed Euclidean space \cite{Faddeev:1987ih}, which is a
noncommutative space with quadratic relations, i.~e. $\theta^{ij}(\hat
{x})=\Theta_{kl}^{ij}\hspace{0.01in}\hat{x}^{k}\hat{x}^{l}$.

The commutation relations for the coordinates of $q$-de\-formed Euclidean
space satisfy the Poincar\'{e}-Birkhoff-Witt property. It says that each
vector space generated by homogeneous polynomials with a fixed degree has the
same dimension as in the commutative case with $\theta^{ij}(\hat{x})=0$. Thus
the nor\-mal-or\-dered monomials of noncommutative coordinates form a basis of
$q$-de\-formed Euclidean space. For this reason, we can associate the
noncommutative algebra of $q$-de\-formed Euclidean space with a commutative
coordinate algebra by using the star-prod\-uct formalism \cite{Moyal:1949sk}.

The star-prod\-uct formalism enables us to construct a $q$-de\-formed version of
mathematical analysis \cite{Carnovale:1999,Wachter:2007A}. In
Ref.~\cite{Wachter:2019A}, we have discussed $q$-de\-formed momentum
eigenfunctions within the framework of this $q$-de\-formed analysis. We have
shown in Ref.~\cite{Wachter:2020A} that the time evolution operator of a
quantum system in $q$-de\-formed Euclidean space is of the same form as in the
undeformed case. In this article, we are going to apply our findings to a
nonrelativistic particle in $q$-de\-formed Euclidean space.

First, we give $q$-analogs for the Hamilton operator of a free,
nonrelativistic particle. Next, we construct $q$-de\-formed plane wave
solutions to the corresponding Schr\"{o}\-dinger equations. We also show that
the $q$-de\-formed plane waves form a complete orthonormal set of functions.
This fact enables us to write down $q$-de\-formed versions of the propagator
for a nonrelativistic particle. Finally, we show how to calculate expectation
values of position or momentum with the solutions of our $q$-de\-formed
Schr\"{o}\-din\-ger equations.

\section{Preliminaries}

\subsection{Star-products\label{KapQuaZeiEle}}

The three-di\-men\-sion\-al $q$-de\-formed Euclidean space $\mathbb{R}_{q}^{3}$ has
the generators $X^{+}$, $X^{3}$, and $X^{-}$, subject to the following
commutation relations \cite{Lorek:1997eh}:%
\begin{align}
X^{3}X^{+}  &  =q^{2}X^{+}X^{3},\nonumber\\
X^{3}X^{-}  &  =q^{-2}X^{-}X^{3},\nonumber\\
X^{-}X^{+}  &  =X^{+}X^{-}+(q-q^{-1})\hspace{0.01in}X^{3}X^{3}.
\label{RelQuaEukDre}%
\end{align}
We can extend the algebra of $\mathbb{R}_{q}^{3}$ by a time element $X^{0}$,
which commutes with the generators $X^{+}$, $X^{3}$, and $X^{-}$
\cite{Wachter:2020A}:%
\begin{equation}
X^{0}X^{A}=X^{A}X^{0},\text{\qquad}A\in\{+,3,-\}. \label{ZusRelExtDreEukQUa}%
\end{equation}
In the following, we refer to the algebra spanned by the generators $X^{i}$
with $i\in\{0,+,3,-\}$ as $\mathbb{R}_{q}^{3,t}$.

There is a $q$-analog of the three-di\-men\-sion\-al Euclidean metric $g^{AB}$ with
its inverse $g_{AB}$ \cite{Lorek:1997eh} (rows and columns are arranged in the
order $+,3,-$):%
\begin{equation}
g_{AB}=g^{AB}=\left(
\begin{array}
[c]{ccc}%
0 & 0 & -\hspace{0.01in}q\\
0 & 1 & 0\\
-\hspace{0.01in}q^{-1} & 0 & 0
\end{array}
\right)  .
\end{equation}
We can use the $q$-de\-formed metric to raise and lower indices:%
\begin{equation}
X_{A}=g_{AB}\hspace{0.01in}X^{B},\qquad X^{A}=g^{AB}X_{B}. \label{HebSenInd}%
\end{equation}

The algebra $\mathbb{R}_{q}^{3,t}$ has a semilinear, involutive, and
anti-multiplicative mapping, which we call \textit{quantum space conjugation}.
If we indicate conjugate elements of a quantum space by a bar,\footnote{A bar
over a complex number indicates complex conjugation.} we can write the
properties of quantum space conjugation as follows ($\alpha,\beta\in
\mathbb{C}$ and $u,v\in\mathbb{R}_{q}^{3,t}$):%
\begin{equation}
\overline{\alpha\,u+\beta\,v}=\overline{\alpha}\,\overline{u}+\overline{\beta
}\,\overline{v},\quad\overline{\overline{u}}=u,\quad\overline{u\,v}%
=\overline{v}\,\overline{u}.
\end{equation}
The conjugation for $\mathbb{R}_{q}^{3,t}$ is compatible with the commutation
relations in Eq.~(\ref{RelQuaEukDre}) and Eq.~(\ref{ZusRelExtDreEukQUa}) if the
following applies \cite{Wachter:2020A}:%
\begin{equation}
\overline{X^{A}}=X_{A}=g_{AB}\hspace{0.01in}X^{B},\qquad\overline{X^{0}}=X_{0}.
\label{ConSpaKoo}%
\end{equation}

We can only prove a physical theory if it predicts measurement results. The
problem, however, is: How can we associate the elements of the noncommutative
space $\mathbb{R}_{q}^{3,t}$ with real numbers? One solution to this problem
is to introduce a vector space isomorphism between the noncommutative algebra
$\mathbb{R}_{q}^{3,t}$ and a corresponding commutative coordinate algebra
$\mathbb{C}[\hspace{0.01in}x^{+},x^{3},x^{-},t\hspace{0.01in}]$.

We recall that the nor\-mal-or\-dered monomials in the generators $X^{i}$ form a
basis of the algebra $\mathbb{R}_{q}^{3,t}$, i.~e. we can write each element
$F\in$ $\mathbb{R}_{q}^{3,t}$ uniquely as a finite or infinite linear
combination of monomials with a given normal ordering
(\textit{Poincar\'{e}-Birkhoff-Witt property}):%
\begin{equation}
F=\sum\limits_{n_{+},\ldots,\hspace{0.01in}n_{0}}a_{\hspace{0.01in}n_{+}%
\ldots\hspace{0.01in}n_{0}}\,(X^{+})^{n_{+}}(X^{3})^{n_{3}}(X^{-})^{n_{-}%
}(X^{0})^{n_{0}},\quad\quad a_{\hspace{0.01in}n_{+}\ldots\hspace{0.01in}n_{0}%
}\in\mathbb{C}.
\end{equation}
Since the monomials $(x^{+})^{n_{+}}(x^{3})^{n_{3}}(x^{-})^{n_{-}}%
t^{\hspace{0.01in}n_{0}}$ with $n_{+},\ldots,n_{0}\in\mathbb{N}_{0}$ form a
basis of the commutative algebra $\mathbb{C}[\hspace{0.01in}x^{+},x^{3}%
,x^{-},t\hspace{0.01in}]$, we can define a vector space isomorphism%
\begin{equation}
\mathcal{W}:\mathbb{C}[\hspace{0.01in}x^{+},x^{3},x^{-},t\hspace{0.01in}]\rightarrow
\mathbb{R}_{q}^{3,t} \label{VecRauIsoInv}%
\end{equation}
with%
\begin{equation}
\mathcal{W}\left(  (x^{+})^{n_{+}}(x^{3})^{n_{3}}(x^{-})^{n_{-}}%
t^{\hspace{0.01in}n_{0}}\right)  =(X^{+})^{n_{+}}(X^{3})^{n_{3}}(X^{-}%
)^{n_{-}}(X^{0})^{n_{0}}. \label{StePro0}%
\end{equation}
In general, we have%
\begin{equation}
\mathbb{C}[\hspace{0.01in}x^{+},x^{3},x^{-},t\hspace{0.01in}]\ni f\mapsto F\in\mathbb{R}%
_{q}^{3,t},
\end{equation}
where%
\begin{align}
f  &  =\sum\limits_{n_{+},\ldots,\hspace{0.01in}n_{0}}a_{\hspace{0.01in}%
n_{+}\ldots\hspace{0.01in}n_{0}}\,(x^{+})^{n_{+}}(x^{3})^{n_{3}}(x^{-}%
)^{n_{-}}t^{\hspace{0.01in}n_{0}},\nonumber\\
F  &  =\sum\limits_{n_{+},\ldots,\hspace{0.01in}n_{0}}a_{\hspace{0.01in}%
n_{+}\ldots\hspace{0.01in}n_{0}}\,(X^{+})^{n_{+}}(X^{3})^{n_{3}}(X^{-}%
)^{n_{-}}(X^{0})^{n_{0}}. \label{AusFfNorOrd}%
\end{align}
The vector space isomorphism $\mathcal{W}$ is nothing else but the 
\textit{Moyal-Weyl mapping}, which gives an operator $F$ to a complex valued
function $f$
\cite{Bayen:1977ha,1997q.alg.....9040K,Madore:2000en,Moyal:1949sk}.

We can extend this vector space isomorphism to an algebra isomorphism if we
introduce a new product on the commutative coordinate algebra. This so-called
\textit{star-prod\-uct }symbolized by $\circledast$ satisfies the following
homomorphism condition:%
\begin{equation}
\mathcal{W}\left(  f\circledast g\right)  =\mathcal{W}\left(  f\right)
\cdot\mathcal{W}\left(  \hspace{0.01in}g\right)  . \label{HomBedWeyAbb}%
\end{equation}
Since the Mo\-yal-Weyl mapping is invertible, we can write the star-prod\-uct as
follows:%
\begin{equation}
f\circledast g=\mathcal{W}^{\hspace{0.01in}-1}\big (\,\mathcal{W}\left(
f\right)  \cdot\mathcal{W}\left(  \hspace{0.01in}g\right)  \big ).
\label{ForStePro}%
\end{equation}

To get explicit formulas for calculating star-prod\-ucts, we first have to write
a noncommutative product of two nor\-mal-or\-dered monomials as a linear
combination of nor\-mal-or\-dered monomials again (see Ref.~\cite{Wachter:2002A}
for details):%
\begin{equation}
(X^{+})^{n_{+}}\ldots\hspace{0.01in}(X^{0})^{n_{0}}\cdot(X^{+})^{m_{+}}%
\ldots\hspace{0.01in}(X^{0})^{m_{0}}=\sum_{\underline{k}\hspace{0.01in}=\hspace{0.01in}%
0}B_{\underline{k}}^{\hspace{0.01in}\underline{n},\underline{m}}\,(X^{+})^{k_{+}}\ldots\hspace{0.01in}%
(X^{0})^{k_{0}}. \label{EntProMon}%
\end{equation}
We achieve this by using the commutation relations for the noncommutative
coordinates [cf. Eq.~(\ref{RelQuaEukDre})]. From the concrete form of the
series expansion in\ Eq.~(\ref{EntProMon}), we can finally read off a formula to
calculate the star-prod\-uct of two power series in commutative space-time
coordinates ($\lambda=q-q^{-1}$):\
\begin{gather}
f(\mathbf{x},t)\circledast g(\mathbf{x},t)=\nonumber\\
\sum_{k\hspace{0.01in}=\hspace{0.01in}0}^{\infty}\lambda^{k}\hspace
{0.01in}\frac{(x^{3})^{2k}}{[[k]]_{q^{4}}!}\,q^{2(\hat{n}_{3}\hspace
{0.01in}\hat{n}_{+}^{\prime}+\,\hat{n}_{-}\hat{n}_{3}^{\prime})}%
D_{q^{4},\hspace{0.01in}x^{-}}^{k}f(\mathbf{x},t)\,D_{q^{4},\hspace
{0.01in}x^{\prime+}}^{k}g(\mathbf{x}^{\prime},t)\big|_{x^{\prime}%
\rightarrow\hspace{0.01in}x}. \label{StaProForExp}%
\end{gather}
The argument $\mathbf{x}$
indicates a dependence on the spatial coordinates $x^{+}$, $x^{3}$, and
$x^{-}$.
Note that the expression\ above depends on the operators%
\begin{equation}
\hat{n}_{A}=x^{A}\frac{\partial}{\partial x^{A}}%
\end{equation}
and the so-called Jackson derivatives \cite{Jackson:1910yd}:%
\begin{equation}
D_{q^{k},\hspace{0.01in}x}\,f=\frac{f(q^{k}x)-f(x)}{q^{k}x-x}.
\end{equation}
Moreover, the $q$-numbers are given by%
\begin{equation}
\lbrack\lbrack a]]_{q}=\frac{1-q^{a}}{1-q},
\end{equation}
and the $q$-factorials are defined in complete analogy to the undeformed case:%
\begin{equation}
\lbrack\lbrack\hspace{0.01in}n]]_{q}!=[[1]]_{q}\hspace{0.01in}[[2]]_{q}%
\ldots\lbrack\lbrack\hspace{0.01in}n-1]]_{q}\hspace{0.01in}[[\hspace
{0.01in}n]]_{q},\qquad\lbrack\lbrack0]]_{q}!=1.
\end{equation}

The algebra isomorphism $\mathcal{W}^{-1}$ also enables us to carry over the
conjugation for the quantum space algebra $
\mathbb{R}_{q}^{3,t}$ to the commutative
coordinate algebra $\mathbb{C}[\hspace{0.01in}x^{+},x^{3},x^{-},t\hspace{0.01in}]$. In other words, the mapping $\mathcal{W}%
^{\hspace{0.01in}-1}$ is a $\ast$-al\-ge\-bra homomorphism:%
\begin{equation}
\mathcal{W}(\hspace{0.01in}\overline{f}\hspace{0.01in})=\overline
{\mathcal{W}(f)}\qquad\Leftrightarrow\text{\qquad}\overline{f}=\mathcal{W}%
^{-1}\big (\hspace{0.01in}\overline{\mathcal{W}(f)}\hspace{0.01in}\big ).
\label{ConAlgIso}%
\end{equation}
This relationship implies the following property for the star-pro\-duct:%
\begin{equation}
\overline{f\circledast g}=\overline{g}\circledast\overline{f}.
\label{KonEigSteProFkt}%
\end{equation}

With $\bar{f}$, we designate the power series obtained from $f$ by quantum space 
conjugation. If $\bar{a}_{n_{+},n_{3},n_{-},n_{0}}$ stands for the complex
conjugate of $a_{n_{+},n_{3},n_{-},n_{0}}$, Eqs.~(\ref{ConSpaKoo})\ and (\ref{ConAlgIso}) yield that  $\bar{f}$ takes the following form
\cite{Wachter:2007A,Wachter:2020A}:%
\begin{align}
\overline{f(\mathbf{x},t)}  &  =\sum\nolimits_{\underline{n}}\bar{a}_{\hspace{0.01in}n_{+},n_{3}%
,n_{-},n_{0}}\,(-\hspace{0.01in}q\hspace{0.01in}x^{-})^{n_{+}}(\hspace
{0.01in}x^{3})^{n_{3}}(-\hspace{0.01in}q^{-1}x^{+})^{n_{-}}t^{n_{0}%
}\nonumber\\
&  =\sum\nolimits_{\underline{n}}(-\hspace{0.01in}q)^{n_{-}-\hspace{0.01in}n_{+}}%
\hspace{0.01in}\bar{a}_{\hspace{0.01in}n_{-},n_{3},n_{+},n_{0}}\,(\hspace{0.01in}%
x^{+})^{n_{+}}(\hspace{0.01in}x^{3})^{n_{3}}(\hspace{0.01in}x^{-})^{n_{-}%
}t^{n_{0}}\nonumber\\
&  =\bar{f}(\mathbf{x},t). \label{KonPotReiKom}%
\end{align}

\subsection{Partial derivatives and integrals\label{KapParDer}}

There are partial derivatives for $q$-de\-formed space-time coordinates
\cite{CarowWatamura:1990zp,Wess:1990vh}. These partial derivatives again form
a quantum space with the same algebraic structure as that of the
$q$-de\-formed space-time coordinates. Thus, the $q$-de\-formed partial
derivatives $\partial_{i}$ satisfy the same commutation relations as the
covariant coordinate generators $X_{i}$:%
\begin{gather}
\partial_{0}\hspace{0.01in}\partial_{+}=\hspace{0.01in}\partial_{+}%
\hspace{0.01in}\partial_{0},\quad\partial_{0}\hspace{0.01in}\partial
_{-}=\hspace{0.01in}\partial_{-}\hspace{0.01in}\partial_{0},\quad\partial
_{0}\hspace{0.01in}\partial_{3}=\partial_{3}\hspace{0.01in}\partial
_{0},\nonumber\\
\partial_{+}\hspace{0.01in}\partial_{3}=q^{2}\partial_{3}\hspace
{0.01in}\partial_{+},\quad\partial_{3}\hspace{0.01in}\partial_{-}%
=\hspace{0.01in}q^{2}\partial_{-}\hspace{0.01in}\partial_{3},\nonumber\\
\partial_{+}\hspace{0.01in}\partial_{-}-\partial_{-}\hspace{0.01in}%
\partial_{+}=\hspace{0.01in}\lambda\hspace{0.01in}\partial_{3}\hspace
{0.01in}\partial_{3}.
\end{gather}
The commutation relations above are invariant under conjugation if the
derivatives show the following conjugation properties:\footnote{The indices of
partial derivatives are raised and lowered in the same way as those of
coordinates [see Eq.~(\ref{HebSenInd}) in Chap.~\ref{KapQuaZeiEle}].}%
\begin{equation}
\overline{\partial_{A}}=-\hspace{0.01in}\partial^{A}=-g^{AB}\partial
_{B},\qquad\overline{\partial_{0}}=-\hspace{0.01in}\partial^{0}=-\hspace
{0.01in}\partial_{0}. \label{KonAbl}%
\end{equation}

There are two ways of commuting $q$-de\-formed partial derivatives with
$q$-de\-formed space-time coordinates. One is given by the following
$q$-de\-formed Leibniz rules
\cite{CarowWatamura:1990zp,Wess:1990vh,Wachter:2020A}:%
\begin{align}
\partial_{B}X^{A}  &  =\delta_{B}^{A}+q^{4}\hat{R}{^{AC}}_{BD}\,X^{D}%
\partial_{C},\nonumber\\
\partial_{A}X^{0}  &  =X^{0}\hspace{0.01in}\partial_{A},\nonumber\\
\partial_{0}\hspace{0.01in}X^{A}  &  =X^{A}\hspace{0.01in}\partial
_{0},\nonumber\\
\partial_{0}\hspace{0.01in}X^{0}  &  =1+X^{0}\hspace{0.01in}\partial_{0}.
\label{DifKalExtEukQuaDreUnk}%
\end{align}
Note that $\hat{R}{^{AC}}_{BD}$ denotes the vector representation of the
R-matrix for the three-di\-men\-sion\-al $q$-de\-formed Euclidean space.

By conjugation, we can obtain the Leibniz rules for another differential
calculus from the identities in Eq.~(\ref{DifKalExtEukQuaDreUnk}). Introducing
$\hat{\partial}_{A}=q^{6}\partial_{A}$ and $\hat{\partial}_{0}=\partial_{0}$,
we can write the Leibniz rules of this second differential calculus in the
following form:%
\begin{align}
\hat{\partial}_{B}\hspace{0.01in}X^{A}  &  =\delta_{B}^{A}+q^{-4}(\hat{R}%
^{-1}){^{AC}}_{BD}\,X^{D}\hat{\partial}_{C},\nonumber\\
\hat{\partial}_{A}\hspace{0.01in}X^{0}  &  =X^{0}\hspace{0.01in}\hat{\partial
}_{A},\nonumber\\
\hat{\partial}_{0}\hspace{0.01in}X^{A}  &  =X^{A}\hspace{0.01in}\hat{\partial
}_{0},\nonumber\\
\hat{\partial}_{0}\hspace{0.01in}X^{0}  &  =1+X^{0}\hspace{0.01in}%
\hat{\partial}_{0}. \label{DifKalExtEukQuaDreKon}%
\end{align}

Using the Leibniz rules in Eq.$~$(\ref{DifKalExtEukQuaDreUnk}) or
Eq.$~$(\ref{DifKalExtEukQuaDreKon}), we can calculate how partial derivatives
act on nor\-mal-or\-dered monomials of noncommutative coordinates. We can
carry over these actions to commutative coordinate monomials with the help of
the Mo\-yal-Weyl mapping:%
\begin{equation}
\partial^{i}\triangleright(x^{+})^{n_{+}}(x^{3})^{n_{3}}(x^{-})^{n_{-}%
}t^{\hspace{0.01in}n_{0}}=\mathcal{W}^{\hspace{0.01in}-1}\big (\partial
^{i}\triangleright(X^{+})^{n_{+}}(X^{3})^{n_{3}}(X^{-})^{n_{-}}(X^{0})^{n_{0}%
}\big ).
\end{equation}
Since the Mo\-yal-Weyl mapping is linear, we can apply the action above to
space-time functions that can be written as a power series:%
\begin{equation}
\partial^{i}\triangleright f(\mathbf{x},t)=\mathcal{W}^{\hspace{0.01in}%
-1}\big (\partial^{i}\triangleright\mathcal{W}(f(\mathbf{x},t))\big ).
\end{equation}

If we use the ordering given in Eq.~(\ref{StePro0}) of the previous chapter,
the Leibniz rules in Eq.~(\ref{DifKalExtEukQuaDreUnk})\ lead to the following
operator representations \cite{Bauer:2003}:%
\begin{align}
\partial_{+}\triangleright f(\mathbf{x},t)  &  =D_{q^{4},\hspace{0.01in}x^{+}%
}f(\mathbf{x},t),\nonumber\\
\partial_{3}\triangleright f(\mathbf{x},t)  &  =D_{q^{2},\hspace{0.01in}x^{3}%
}f(q^{2}x^{+},x^{3},x^{-},t),\nonumber\\
\partial_{-}\triangleright f(\mathbf{x},t)  &  =D_{q^{4},\hspace{0.01in}x^{-}%
}f(x^{+},q^{2}x^{3},x^{-},t)+\lambda\hspace{0.01in}x^{+}D_{q^{2}%
,\hspace{0.01in}x^{3}}^{2}f(\mathbf{x},t). \label{UnkOpeDarAbl}%
\end{align}
The derivative $\partial_{0}$, however, is represented on the commutative
space-time algebra by an ordinary partial derivative:%
\begin{equation}
\partial_{0}\triangleright\hspace{-0.01in}f(\mathbf{x},t)=\frac{\partial
f(\mathbf{x},t)}{\partial t}. \label{OpeDarZeiAblExtQuaEuk}%
\end{equation}

Using the Leibniz rules in Eq.$~$(\ref{DifKalExtEukQuaDreKon}), we get operator
representations for the partial derivatives $\hat{\partial}_{i}$. The Leibniz
rules in Eq.$~$(\ref{DifKalExtEukQuaDreUnk}) and Eq.$~$%
(\ref{DifKalExtEukQuaDreKon}) are transformed into each other by the following
substitutions:%
\begin{gather}
q\rightarrow q^{-1},\quad X^{-}\rightarrow X^{+},\quad X^{+}\rightarrow
X^{-},\nonumber\\
\partial^{\hspace{0.01in}+}\rightarrow\hat{\partial}^{\hspace{0.01in}-}%
,\quad\partial^{\hspace{0.01in}-}\rightarrow\hat{\partial}^{\hspace{0.01in}%
+},\quad\partial^{\hspace{0.01in}3}\rightarrow\hat{\partial}^{\hspace
{0.01in}3},\quad\partial^{\hspace{0.01in}0}\rightarrow\hat{\partial}%
^{\hspace{0.01in}0}. \label{UebRegGedUngAblDreQua}%
\end{gather}
For this reason, we obtain the operator representations of the partial
derivatives $\hat{\partial}_{A}$ from those of the partial derivatives
$\partial_{A}$ [cf. Eq.~(\ref{UnkOpeDarAbl})] if we replace $q$ by $q^{-1}$
and exchange the indices $+$ and $-$:%
\begin{align}
\hat{\partial}_{-}\,\bar{\triangleright}\,f(\mathbf{x},t)  &  =D_{q^{-4}%
,\hspace{0.01in}x^{-}}f(\mathbf{x},t),\nonumber\\
\hat{\partial}_{3}\,\bar{\triangleright}\,f(\mathbf{x},t)  &  =D_{q^{-2}%
,\hspace{0.01in}x^{3}}f(q^{-2}x^{-},x^{3},x^{+},t),\nonumber\\
\hat{\partial}_{+}\,\bar{\triangleright}\,f(\mathbf{x},t)  &  =D_{q^{-4}%
,\hspace{0.01in}x^{+}}f(x^{-},q^{-2}x^{3},x^{+},t)-\lambda\hspace{0.01in}%
x^{-}D_{q^{-2},\hspace{0.01in}x^{3}}^{2}f(\mathbf{x},t). \label{KonOpeDarAbl}%
\end{align}
Once again, $\hat{\partial}_{0}$ is represented on the commutative space-time
algebra by an ordinary partial derivative:%
\begin{equation}
\hat{\partial}_{0}\,\bar{\triangleright}\,f(\mathbf{x},t)=\frac{\partial
f(\mathbf{x},t)}{\partial t}. \label{OpeDarZeiAblExtQuaEukKon}%
\end{equation}
Due to the substitutions given in\ Eq.~(\ref{UebRegGedUngAblDreQua}), the actions in
Eqs.~(\ref{KonOpeDarAbl}) and (\ref{OpeDarZeiAblExtQuaEukKon}) refer to
nor\-mal-or\-dered monomials different from those in Eq.~(\ref{StePro0}) of the
previous chapter:%
\begin{equation}
\widetilde{\mathcal{W}}\left(  t^{\hspace{0.01in}n_{0}}(x^{+})^{n_{+}}%
(x^{3})^{n_{3}}(x^{-})^{n_{-}}\right)  =(X^{0})^{n_{0}}(X^{-})^{n_{-}}%
(X^{3})^{n_{3}}(X^{+})^{n_{+}}. \label{UmNor}%
\end{equation}

We should not forget that we can also commute
$q$-de\-formed partial derivatives from the \textit{right} side of a
nor\-mal-or\-dered monomial to the left side by using the Leibniz rules. This way,
we get\ the so-called \textit{right}-re\-pre\-sen\-ta\-tions of partial
derivatives, for which we write $f\,\bar{\triangleleft}\,\partial^{i}$ or
$f\triangleleft\hat{\partial}^{i}$. Note that the operation of conjugation
transforms left actions of partial derivatives into right actions and vice
versa \cite{Bauer:2003}:%
\begin{align}
\overline{\partial^{i}\triangleright f}  &  =-\bar{f}\,\bar{\triangleleft
}\,\partial_{i}, & \overline{f\,\bar{\triangleleft}\,\partial^{i}}  &
=-\hspace{0.01in}\partial_{i}\triangleright\bar{f},\nonumber\\
\overline{\hat{\partial}^{i}\,\bar{\triangleright}\,f}  &  =-\bar
{f}\triangleleft\hat{\partial}_{i}, & \overline{f\triangleleft\hat{\partial
}^{i}}  &  =-\hspace{0.01in}\hat{\partial}_{i}\,\bar{\triangleright}\,\bar{f}.
\label{RegConAbl}%
\end{align}

In general, the operator representations in Eqs.~(\ref{UnkOpeDarAbl}) and
(\ref{KonOpeDarAbl}) consist of two terms, which we call $\partial
_{\operatorname*{cla}}^{A}$ and $\partial_{\operatorname*{cor}}^{A}$:%
\begin{equation}
\partial^{A}\triangleright F=\left(  \partial_{\operatorname*{cla}}%
^{A}+\partial_{\operatorname*{cor}}^{A}\right)  \triangleright F.
\end{equation}
In the undeformed limit $q\rightarrow1$, $\partial_{\operatorname*{cla}}^{A}$
becomes an ordinary partial derivative, and $\partial_{\operatorname*{cor}%
}^{A}$ disappears. We get a solution to the difference equation $\partial
^{A}\triangleright F=f$ with given $f$ by using the following formula
\cite{Wachter:2004A}:%
\begin{align}
F  &  =(\partial^{A})^{-1}\triangleright f=\left(  \partial
_{\operatorname*{cla}}^{A}+\partial_{\operatorname*{cor}}^{A}\right)
^{-1}\triangleright f\nonumber\\
&  =\sum_{k\hspace{0.01in}=\hspace{0.01in}0}^{\infty}\left[  -(\partial
_{\operatorname*{cla}}^{A})^{-1}\partial_{\operatorname*{cor}}^{A}\right]
^{k}(\partial_{\operatorname*{cla}}^{A})^{-1}\triangleright f.
\end{align}
Applying the above formula to the operator representations in
Eq.~(\ref{UnkOpeDarAbl}), we get%
\begin{align}
(\partial_{+})^{-1}\triangleright f(\mathbf{x},t)  &  =D_{q^{4},\hspace
{0.01in}x^{+}}^{-1}f(\mathbf{x},t),\nonumber\\
(\partial_{3})^{-1}\triangleright f(\mathbf{x},t)  &  =D_{q^{2},\hspace
{0.01in}x^{3}}^{-1}f(q^{-2}x^{+},x^{3},x^{-},t), \label{InvParAbl1}%
\end{align}
and%
\begin{gather}
(\partial_{-})^{-1}\triangleright f(\mathbf{x},t)=\nonumber\\
=\sum_{k\hspace{0.01in}=\hspace{0.01in}0}^{\infty}q^{2k\left(  k\hspace
{0.01in}+1\right)  }\left(  -\lambda\,x^{+}D_{q^{4},\hspace{0.01in}x^{-}}%
^{-1}D_{q^{2},\hspace{0.01in}x^{3}}^{2}\right)  ^{k}D_{q^{4},\hspace
{0.01in}x^{-}}^{-1}f(x^{+},q^{-2\left(  k\hspace{0.01in}+1\right)  }%
x^{3},x^{-},t). \label{InvParAbl2}%
\end{gather}
Note that $D_{q,\hspace{0.01in}x}^{-1}$ stands for a Jackson integral with $x$
being the variable of integration \cite{Jackson:1908}. The explicit form of
this Jackson integral depends on its limits of integration and the value for
the deformation parameter $q$. If $x>0$ and $q>1$, for example, the following
applies:%
\begin{equation}
\int_{0}^{\hspace{0.01in}x}\text{d}_{q}z\hspace{0.01in}f(z)=(q-1)\hspace
{0.01in}x\sum_{j=1}^{\infty}q^{-j}f(q^{-j}x).
\end{equation}
Finally, the integral for the time coordinate is an ordinary integral since
$\partial_{0}$ acts on the commutative space-time algebra like an ordinary
partial derivative [cf. Eq.~(\ref{OpeDarZeiAblExtQuaEuk})]:%
\begin{equation}
(\partial_{0})^{-1}\triangleright f(\mathbf{x},t)\hspace{0.01in}=\int
\text{d}t\,f(\mathbf{x},t).
\end{equation}

The above considerations also apply to the partial derivatives with a hat.
However, we can obtain the representations of $\hat{\partial}_{i}$ from those
of the derivatives $\partial_{i}$ if we replace $q$ with $q^{-1}$ and exchange
the indices $+$ and $-$. Applying these substitutions to the expressions in
Eqs.~(\ref{InvParAbl1}) and (\ref{InvParAbl2}), we immediately get the
corresponding results for the partial derivatives $\hat{\partial}_{i}$.

By successively applying the integral operators given in
Eqs.~(\ref{InvParAbl1}) and (\ref{InvParAbl2}), we can explain an integration
over all space \cite{Wachter:2004A,Wachter:2007A}:%
\begin{equation}
\int_{-\infty}^{+\infty}\text{d}_{q}^{3}x\,f(x^{+},x^{3},x^{-})=(\partial
_{-})^{-1}\big |_{-\infty}^{+\infty}\,(\partial_{3})^{-1}\big |_{-\infty
}^{+\infty}\,(\partial_{+})^{-1}\big |_{-\infty}^{+\infty}\triangleright f.
\end{equation}
On the right-hand side of the above relation,the different integral operators
can be simplified to Jackson integrals \cite{Wachter:2004A,Jambor:2004ph}:%
\begin{equation}
\int_{-\infty}^{+\infty}\text{d}_{q}^{3}x\,f(\mathbf{x})=D_{q^{2}%
,\hspace{0.01in}x^{-}}^{-1}\big |_{-\infty}^{+\infty}\,D_{q,x^{3}}%
^{-1}\big |_{-\infty}^{+\infty}\,D_{q^{2},\hspace{0.01in}x^{+}}^{-1}%
\big |_{-\infty}^{+\infty}\,f(\mathbf{x}).
\end{equation}
Note that the Jackson integrals in the formula above refer to a smaller
$q$-lattice. Using such a smaller $q$-lattice ensures that our integral over
all space is a scalar with trivial braiding properties \cite{Kempf:1994yd}.

The $q$-integral over all space shows some significant features
\cite{Wachter:2007A,Jambor:2004ph}. In this respect, $q$-de\-formed versions of
\textit{Stokes' theorem} apply:%
\begin{align}
\int_{-\infty}^{+\infty}\text{d}_{q}^{3}x\,\partial^{A}\triangleright f  &
=\int_{-\infty}^{+\infty}\text{d}_{q}^{3}x\,f\,\bar{\triangleleft}%
\,\partial^{A}=0,\nonumber\\
\int_{-\infty}^{+\infty}\text{d}_{q}^{3}x\,\hat{\partial}^{A}\,\bar
{\triangleright}\,f  &  =\int_{-\infty}^{+\infty}\text{d}_{q}^{3}%
x\,f\triangleleft\hat{\partial}^{A}=0.
\end{align}
The $q$-de\-formed Stokes' theorem also implies rules for integration by
parts:%
\begin{align}
\int_{-\infty}^{+\infty}\text{d}_{q}^{3}x\,f\circledast(\partial
^{A}\triangleright g)  &  =\int_{-\infty}^{+\infty}\text{d}_{q}^{3}%
x\,(f\triangleleft\partial^{A})\circledast g,\nonumber\\
\int_{-\infty}^{+\infty}\text{d}_{q}^{3}x\,f\circledast(\hat{\partial}%
^{A}\,\bar{\triangleright}\,g)  &  =\int_{-\infty}^{+\infty}\text{d}_{q}%
^{3}x\,(f\,\bar{\triangleleft}\,\hat{\partial}^{A})\circledast g.
\label{PatIntUneRaumInt}%
\end{align}
Finally, we mention that the $q$-integral over all space behaves as follows
under quantum space conjugation:%
\begin{equation}
\overline{\int_{-\infty}^{+\infty}\text{d}_{q}^{3}x\,f}=\int_{-\infty
}^{+\infty}\text{d}_{q}^{3}x\,\bar{f}. \label{KonEigVolInt}%
\end{equation}

\subsection{Exponentials and Translations\label{KapExp}}

\begin{figure}
[ptb]
\begin{center}
\centerline{\psfig{figure=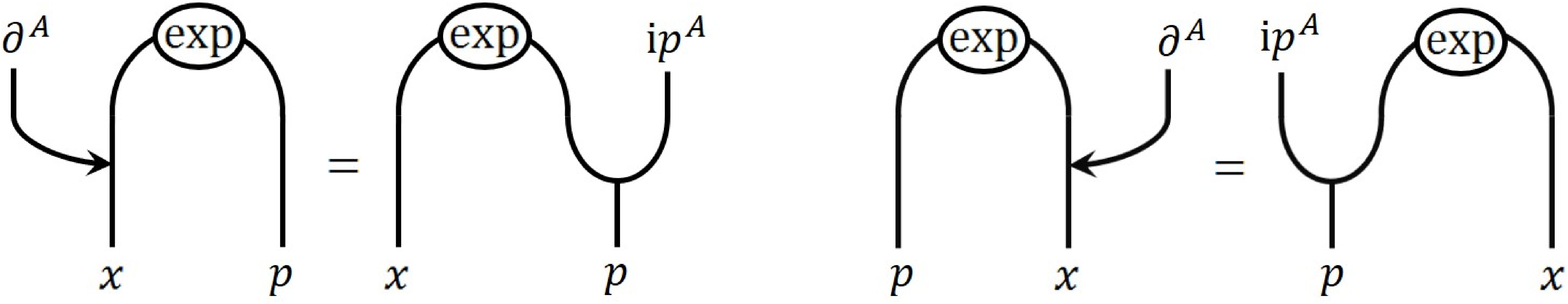,width=4.555in}}%
\caption{Eigenvalue equations of $q$-exponentials.}%
\label{Fig1}%
\end{center}
\end{figure}

A $q$-de\-formed exponential is an eigenfunction of each partial derivative of
a given $q$-de\-formed quantum space
\cite{Majid:1993ud,Schirrmacher:1995,Wachter:2004ExpA}. In the following, we
consider $q$-de\-formed exponentials that are eigenfunctions for left actions
or right actions of partial derivatives:%
\begin{align}
\text{i}^{-1}\partial^{A}\triangleright\exp_{q}(\mathbf{x}|\text{i}\mathbf{p})
&  =\exp_{q}(\mathbf{x}|\text{i}\mathbf{p})\circledast p^{A},\nonumber\\
\exp_{q}(\text{i}^{-1}\mathbf{p}|\hspace{0.01in}\mathbf{x})\,\bar
{\triangleleft}\,\partial^{A}\text{i}^{-1} &  =p^{A}\circledast\exp
_{q}(\text{i}^{-1}\mathbf{p}|\hspace{0.01in}\mathbf{x}).\label{EigGl1N}%
\end{align}
The above eigenvalue equations are shown graphically in Fig.~\ref{Fig1}. The
$q$-ex\-po\-nen\-tials are uniquely defined by their eigenvalue equations and
the following normalization conditions:%
\begin{align}
\exp_{q}(\mathbf{x}|\text{i}\mathbf{p})|_{x\hspace{0.01in}=\hspace{0.01in}0}
&  =\exp_{q}(\mathbf{x}|\text{i}\mathbf{p})|_{p\hspace{0.01in}=\hspace
{0.01in}0}=1,\nonumber\\
\exp_{q}(\text{i}^{-1}\mathbf{p}|\hspace{0.01in}\mathbf{x})|_{x\hspace
{0.01in}=\hspace{0.01in}0} &  =\exp_{q}(\text{i}^{-1}\mathbf{p}|\hspace
{0.01in}\mathbf{x})|_{p\hspace{0.01in}=\hspace{0.01in}0}=1.\label{NorBedExp}%
\end{align}%

Using the operator representation in Eq.~(\ref{UnkOpeDarAbl}) of the last
chapter, we found the following expressions for the $q$-ex\-ponen\-tials of
three-di\-men\-sion\-al Euclidean quantum space \cite{Wachter:2004ExpA}:%
\begin{align}
\exp_{q}(\mathbf{x}|\text{i}\mathbf{p})  &  =\sum_{\underline{n}%
\,=\,0}^{\infty}\frac{(x^{+})^{n_{+}}(x^{3})^{n_{3}}(x^{-})^{n_{-}}%
(\text{i}p_{-})^{n_{-}}(\text{i}p_{3})^{n_{3}}(\text{i}p_{+})^{n_{+}}%
}{[[\hspace{0.01in}n_{+}]]_{q^{4}}!\,[[\hspace{0.01in}n_{3}]]_{q^{2}%
}!\,[[\hspace{0.01in}n_{-}]]_{q^{4}}!},\nonumber\\
\exp_{q}(\text{i}^{-1}\mathbf{p}|\mathbf{x})  &  =\sum_{\underline{n}%
\,=\,0}^{\infty}\frac{(\text{i}^{-1}p^{+})^{n_{+}}(\text{i}^{-1}p^{3})^{n_{3}%
}(\text{i}^{-1}p^{-})^{n_{-}}(x_{-})^{n_{-}}(x_{3})^{n_{3}}(x_{+})^{n_{+}}%
}{[[\hspace{0.01in}n_{+}]]_{q^{4}}!\,[[\hspace{0.01in}n_{3}]]_{q^{2}%
}!\,[[\hspace{0.01in}n_{-}]]_{q^{4}}}. \label{ExpEukExp}%
\end{align}
If we substitute $q$ with $q^{-1}$ in both expressions of Eq.~(\ref{ExpEukExp}%
), we get two more $q$-exponentials, which we designate $\overline{\exp}%
_{q}(x|$i$\mathbf{p})$ and $\overline{\exp}_{q}($i$^{-1}\mathbf{p}|x)$. We
obtain the eigenvalue equations and normalization conditions of these two
$q$-exponentials by applying the following substitutions to
Eqs.~(\ref{EigGl1N}) and (\ref{NorBedExp}):%
\begin{equation}
\exp_{q}\rightarrow\hspace{0.01in}\overline{\exp}_{q},\qquad\triangleright\,\rightarrow
\,\bar{\triangleright},\qquad\bar{\triangleleft}\,\rightarrow\,\triangleleft
,\qquad\partial^{A}\rightarrow\hat{\partial}^{A}. \label{ErsRegQExp}%
\end{equation}

We can use $q$-exponentials to generate $q$-translations \cite{Chryssomalakos:1993zm}. If we replace the
momentum coordinates in the expressions for $q$-exponentials with derivatives,
it applies \cite{Carnovale:1999,Majid:1993ud,Wachter:2007A}%
\begin{align}
\exp_{q}(x|\partial_{y})\triangleright g(\hspace{0.01in}y)  &  =g(x\,\bar
{\oplus}\,y),\nonumber\\
\overline{\exp}_{q}(x|\hat{\partial}_{y})\,\bar{\triangleright}\,g(\hspace
{0.01in}y)  &  =g(x\oplus y), \label{q-TayN}%
\end{align}
and%
\begin{align}
g(\hspace{0.01in}y)\,\bar{\triangleleft}\,\exp_{q}(-\hspace{0.01in}%
\partial_{y}|\hspace{0.01in}x)  &  =g(\hspace{0.01in}y\,\bar{\oplus
}\,x),\nonumber\\
g(\hspace{0.01in}y)\triangleleft\hspace{0.01in}\overline{\exp}_{q}%
(-\hspace{0.01in}\hat{\partial}_{y}|\hspace{0.01in}x)  &  =g(\hspace
{0.01in}y\oplus x). \label{q-TayRecN}%
\end{align}
In the case of the three-di\-men\-sion\-al $q$-de\-formed Euclidean space, for
example, we can get the following formula for calculating $q$-trans\-la\-tions
\cite{Wachter:2004phengl}:%
\begin{align}
f(\mathbf{x}\oplus\mathbf{y})=  &  \sum_{i_{+}=\hspace{0.01in}0}^{\infty}%
\sum_{i_{3}=\hspace{0.01in}0}^{\infty}\sum_{i_{-}=\hspace{0.01in}0}^{\infty
}\sum_{k\hspace{0.01in}=\hspace{0.01in}0}^{i_{3}}\frac{(-q^{-1}\lambda
\lambda_{+})^{k}}{[[2k]]_{q^{-2}}!!}\frac{(x^{-})^{i_{-}}(x^{3})^{i_{3}%
-\hspace{0.01in}k}(x^{+})^{i_{+}+\hspace{0.01in}k}\,(\hspace{0.01in}y^{-})^{k}}%
{[[i_{-}]]_{q^{-4}}!\,[[i_{3}-k]]_{q^{-2}}!\,[[i_{+}]]_{q^{-4}}!}\nonumber\\
&  \qquad\times\big (D_{q^{-4},\hspace{0.01in}y^{-}}^{i_{-}}D_{q^{-2}%
,\hspace{0.01in}y^{3}}^{i_{3}+\hspace{0.01in}k}\hspace{0.01in}D_{q^{-4}%
,\hspace{0.01in}y^{+}}^{i_{+}}f\big )(q^{2(k\hspace{0.01in}-\hspace
{0.01in}i_{3})}y^{-},q^{-2i_{+}}y^{3}).
\end{align}

\begin{figure}
[ptb]
\begin{center}
\centerline{\psfig{figure=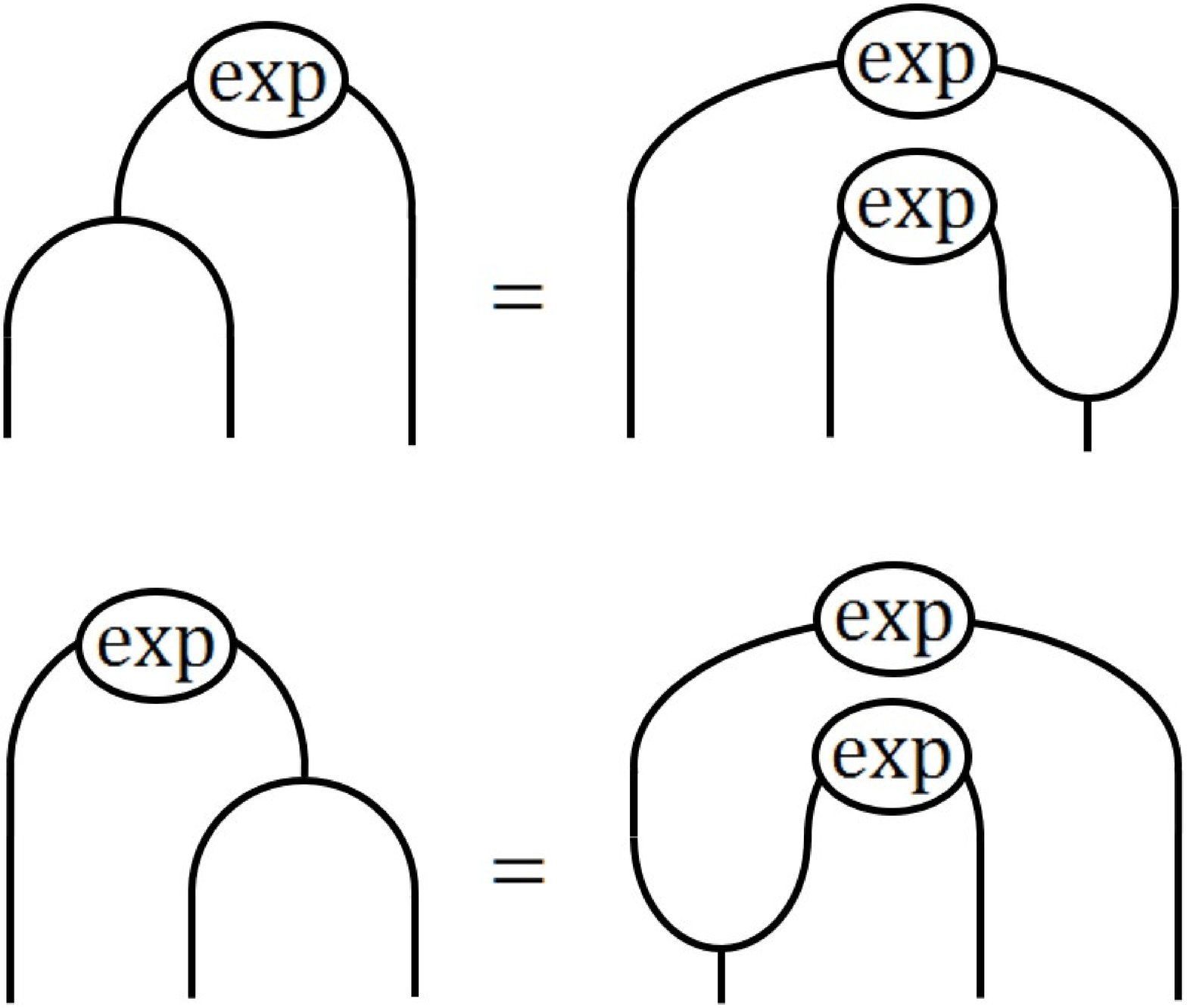,width=1.8827in}}%
\caption{Addition theorems for $q$-exponentials.}%
\label{Fig2}%
\end{center}
\end{figure}

In analogy to the undeformed case, $q$-ex\-ponen\-tials satisfy addition
theorems \cite{Majid:1993ud,Schirrmacher:1995,Wachter:2007A}. Concretely, we
have%
\begin{align}
\exp_{q}(\mathbf{x}\,\bar{\oplus}\,\mathbf{y}|\text{i}\mathbf{p}) &  =\exp
_{q}(\mathbf{x}|\exp_{q}(\hspace{0.01in}\mathbf{y}|\text{i}\mathbf{p}%
)\circledast\text{i}\mathbf{p}),\nonumber\\
\exp_{q}(\text{i}\mathbf{x}|\mathbf{p}\,\bar{\oplus}\,\mathbf{p}^{\prime}) &
=\exp_{q}(\mathbf{x}\circledast\exp_{q}(\mathbf{x}|\hspace{0.01in}%
\text{i}\mathbf{p})|\hspace{0.01in}\text{i}\mathbf{p}^{\prime}%
),\label{AddTheExp}%
\end{align}
and%
\begin{align}
\overline{\exp}_{q}(\mathbf{x}\oplus\mathbf{y}|\text{i}\mathbf{p}) &
=\overline{\exp}_{q}(\mathbf{x}|\overline{\exp}_{q}(\hspace{0.01in}%
\mathbf{y}|\text{i}\mathbf{p})\circledast\text{i}\mathbf{p}),\nonumber\\
\overline{\exp}_{q}(\text{i}\mathbf{x}|\mathbf{p}\oplus\mathbf{p}^{\prime}) &
=\overline{\exp}_{q}(\mathbf{x}\circledast\overline{\exp}_{q}(\mathbf{x}%
|\text{i}\mathbf{p})|\hspace{0.01in}\text{i}\mathbf{p}^{\prime}).
\end{align}
We can obtain further addition theorems from the above identities by
substituting position coordinates with momentum coordinates and vice versa.
For a better understanding of the meaning of the two addition theorems in
Eq.~(\ref{AddTheExp}), we have given their graphic representation in
Fig.~\ref{Fig2}.%

The $q$-de\-formed quantum spaces considered so far are so-called braided Hopf
algebras \cite{Majid:1996kd}. From this point of view, the two versions of
$q$-trans\-lations are nothing else but realizations of two braided
co-pro\-ducts $\underline{\Delta}$ and $\underline{\bar{\Delta}}$ on the
corresponding commutative coordinate algebras \cite{Wachter:2007A}:%
\begin{align}
f(\mathbf{x}\oplus\mathbf{y})  &  =((\mathcal{W}^{\hspace{0.01in}-1}%
\otimes\mathcal{W}^{\hspace{0.01in}-1})\circ\underline{\Delta})(\mathcal{W}%
(f)),\nonumber\\[0.08in]
f(\mathbf{x}\,\bar{\oplus}\,\mathbf{y})  &  =((\mathcal{W}^{\hspace{0.01in}%
-1}\otimes\mathcal{W}^{-1})\circ\underline{\bar{\Delta}})(\mathcal{W}(f)).
\label{KonReaBraCop}%
\end{align}
The braided Hopf algebras have braided antipodes $\underline{S}$ and
$\underline{\bar{S}}$ as well. We can realize these antipodes on the
corresponding commutative algebras, too:%
\begin{align}
f(\ominus\,\mathbf{x})  &  =(\mathcal{W}^{\hspace{0.01in}-1}\circ\underline
{S}\hspace{0.01in})(\mathcal{W}(f)),\nonumber\\
f(\bar{\ominus}\,\mathbf{x})  &  =(\mathcal{W}^{\hspace{0.01in}-1}%
\circ\underline{\bar{S}}\hspace{0.01in})(\mathcal{W}(f)). \label{qInvDef}%
\end{align}
In the following, we refer to the operations in Eq.~(\ref{qInvDef})\ as
$q$\textit{-in\-ver\-sions}. In the case of the $q$-de\-formed Euclidean
space, for example, we have found the following operator representation for
$q$-in\-ver\-sions \cite{Wachter:2004phengl}:%
\begin{align}
\hat{U}^{-1}f(\ominus\,\mathbf{x})=  &  \sum_{i=0}^{\infty}(-\hspace
{0.01in}q\lambda\lambda_{+})^{i}\,\frac{(x^{+}x^{-})^{i}}{[[2i]]_{q^{-2}}%
!!}\,q^{-2\hat{n}_{+}(\hat{n}_{+}+\hspace{0.01in}\hat{n}_{3})-2\hat{n}%
_{-}(\hat{n}_{-}+\hspace{0.01in}\hat{n}_{3})-\hat{n}_{3}\hat{n}_{3}%
}\nonumber\\
&  \qquad\times D_{q^{-2},\hspace{0.01in}x^{3}}^{2i}\,f(-\hspace
{0.01in}q^{2-4i}x^{-},-\hspace{0.01in}q^{1-2i}x^{3},-\hspace{0.01in}%
q^{2-4i}x^{+}).
\end{align}
The operators $\hat{U}$ and $\hat{U}^{-1}$ act on a commutative function
$f(x^{+},x^{3},x^{-})$ as follows:%
\begin{align}
\hat{U}f  &  =\sum_{k\hspace{0.01in}=\hspace{0.01in}0}^{\infty}\left(
-\lambda\right)  ^{k}\frac{(x^{3})^{2k}}{[[k]]_{q^{-4}}!}\,q^{-2\hat{n}%
_{3}(\hat{n}_{+}+\hspace{0.01in}\hat{n}_{-}+\hspace{0.01in}k)}D_{q^{-4}%
,\hspace{0.01in}x^{+}}^{k}D_{q^{-4},\hspace{0.01in}x^{-}}^{k}f,\nonumber\\
\hat{U}^{-1}f  &  =\sum_{k\hspace{0.01in}=\hspace{0.01in}0}^{\infty}%
\lambda^{k}\hspace{0.01in}\frac{(x^{3})^{2k}}{[[k]]_{q^{4}}!}\,q^{2\hat{n}%
_{3}(\hat{n}_{+}+\hspace{0.01in}\hat{n}_{-}+\hspace{0.01in}k)}D_{q^{4}%
,\hspace{0.01in}x^{+}}^{k}D_{q^{4},\hspace{0.01in}x^{-}}^{k}f.
\end{align}

The braided co-products and braided antipodes satisfy the axioms (also see
Ref.~\cite{Majid:1996kd})%
\begin{align}
m\circ(\underline{S}\otimes\operatorname*{id})\circ\underline{\Delta}  &
=m\circ(\operatorname*{id}\otimes\,\underline{S}\hspace{0.01in})\circ
\underline{\Delta}=\underline{\varepsilon},\nonumber\\
m\circ(\underline{\bar{S}}\otimes\operatorname*{id})\circ\underline
{\bar{\Delta}}  &  =m\circ(\operatorname*{id}\otimes\,\underline{\bar{S}%
}\hspace{0.01in})\circ\underline{\bar{\Delta}}=\underline{\bar{\varepsilon}},
\label{HopfVerAnfN}%
\end{align}
and%
\begin{align}
(\operatorname*{id}\otimes\,\underline{\varepsilon})\circ\underline{\Delta}
&  =\operatorname*{id}=(\underline{\varepsilon}\otimes\operatorname*{id}%
)\circ\underline{\Delta},\nonumber\\
(\operatorname*{id}\otimes\,\underline{\bar{\varepsilon}})\circ\underline
{\bar{\Delta}}  &  =\operatorname*{id}=(\underline{\bar{\varepsilon}}%
\otimes\operatorname*{id})\circ\underline{\bar{\Delta}}. \label{HopfAxi2}%
\end{align}
In the identities above, we denote the operation of multiplication on the
braided Hopf algebra by $m$. The co-units $\underline{\varepsilon}%
,\underline{\bar{\varepsilon}}$ of the two braided Hopf structures are both
linear mappings vanishing on the coordinate generators:%
\begin{equation}
\varepsilon(X^{i})=\underline{\bar{\varepsilon}}(X^{i})=0.
\end{equation}
For this reason, we can realize the co-units $\underline{\varepsilon}$ and
$\underline{\bar{\varepsilon}}$ on a commutative coordinate algebra as
follows:%
\begin{equation}
\underline{\varepsilon}(\mathcal{W}(f))=\underline{\bar{\varepsilon}%
}(\mathcal{W}(f))=\left.  f(\mathbf{x})\right\vert _{x\hspace{0.01in}%
=\hspace{0.01in}0}=f(0). \label{ReaVerZopNeuEleKomAlg}%
\end{equation}
Next, we translate the Hopf algebra axioms in Eqs.~(\ref{HopfVerAnfN}) and
(\ref{HopfAxi2}) into corresponding rules for $q$-translations and
$q$-inversions \cite{Wachter:2007A}, i.~e.%
\begin{align}
f((\ominus\,\mathbf{x})\oplus\mathbf{x})  &  =f(\mathbf{x}\oplus
(\ominus\,\mathbf{x}))=f(0),\nonumber\\
f((\bar{\ominus}\,\mathbf{x})\,\bar{\oplus}\,\mathbf{x})  &  =f(\mathbf{x}%
\,\bar{\oplus}\,(\bar{\ominus}\,\mathbf{x}))=f(0), \label{qAddN}%
\end{align}
and%
\begin{align}
f(\mathbf{x}\oplus\mathbf{y})|_{y\hspace{0.01in}=\hspace{0.01in}0}  &
=f(\mathbf{x})=f(\mathbf{y}\oplus\mathbf{x})|_{y\hspace{0.01in}=\hspace
{0.01in}0},\nonumber\\
f(\mathbf{x}\,\bar{\oplus}\,\mathbf{y})|_{y\hspace{0.01in}=\hspace{0.01in}0}
&  =f(\mathbf{x})=f(\mathbf{y}\,\bar{\oplus}\,\mathbf{x})|_{y\hspace
{0.01in}=\hspace{0.01in}0}. \label{qNeuEle}%
\end{align}

\begin{figure}
[ptb]
\centerline{\psfig{figure=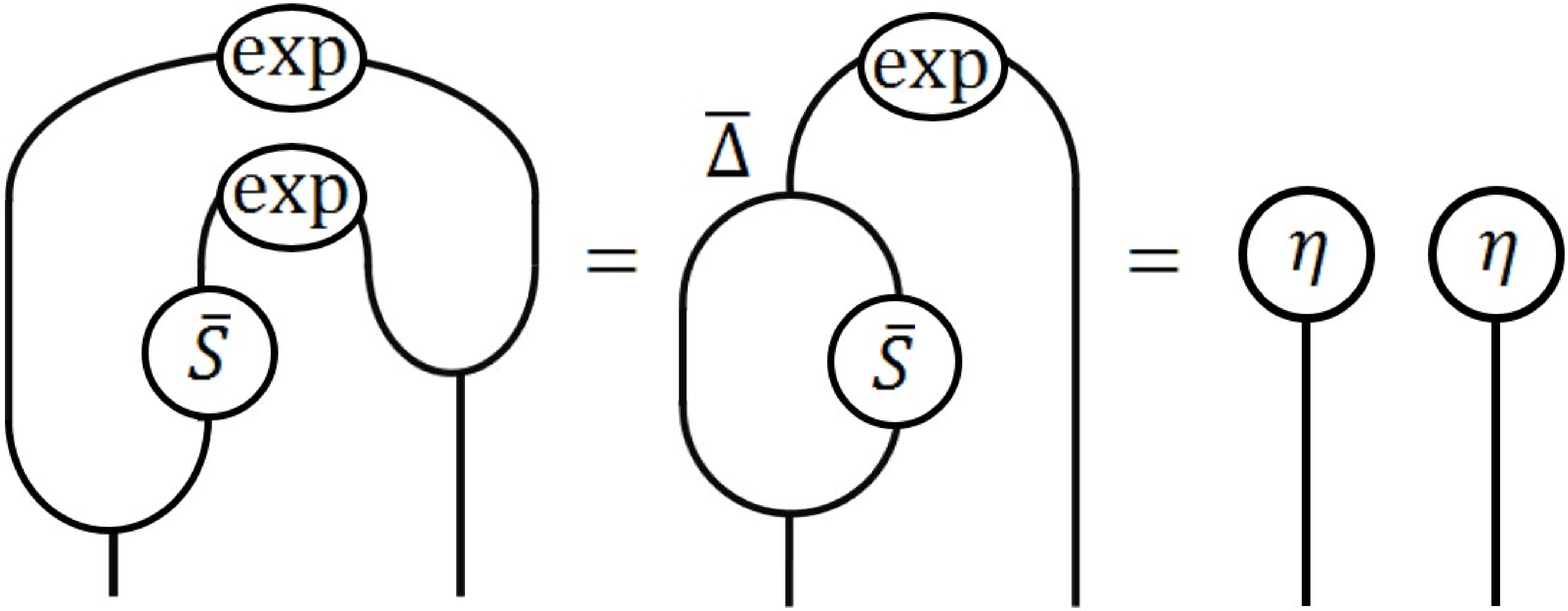,width=2.5754in}}%
\caption{Invertibility of $q$-exponentials.}%
\label{Fig3}
\end{figure}

Using $q$-in\-ver\-sions, we are also able to introduce inverse $q$%
-exponentials:%
\begin{equation}
\exp_{q}(\bar{\ominus}\,\mathbf{x}|\text{i}\mathbf{p})=\exp_{q}(\text{i}%
\mathbf{x}|\text{{}}\bar{\ominus}\,\mathbf{p}).\label{InvExpAlgDefKom}%
\end{equation}
Due to the addition theorems and the normalization conditions of our
$q$-ex\-ponen\-tials, the following applies:%
\begin{equation}
\exp_{q}(\text{i}\mathbf{x}\circledast\exp_{q}(\bar{\ominus}\,\mathbf{x}%
|\hspace{0.01in}\text{i}\mathbf{p})\circledast\mathbf{p})=\exp_{q}%
(\mathbf{x}\,\bar{\oplus}\,(\bar{\ominus}\,\mathbf{x})|\hspace{0.01in}%
\text{i}\mathbf{p})=\exp_{q}(\mathbf{x}|\text{i}\mathbf{p})|_{x=0}=1.
\end{equation}
For a better understanding of these identities, we have given their graphic
representation in Fig.~\ref{Fig3}. You find some explanations of this
sort of graphical calculations in Ref.~\cite{Majid:2002kd}.
The conjugate $q$-exponentials $\overline{\exp}_{q}$ are subject to similar
rules obtained from the above identities by using the following substitutions:%
\begin{equation}
\exp_{q}\rightarrow\overline{\exp}_{q},\qquad\bar{\oplus}\,\rightarrow
\,\oplus,\qquad\bar{\ominus}\,\rightarrow\,\ominus.
\end{equation}

\begin{figure}
[ptb]
\centerline{\psfig{figure=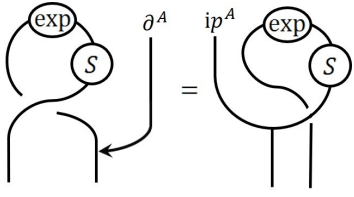,width=1.817in}}%
\caption{Eigenvalue equation of twisted $q$-exponential.}%
\label{Fig4}%
\end{figure}

Next, we describe another way of obtaining $q$-ex\-ponentials. We exchange the
two tensor factors of a $q$-ex\-ponential using the inverse of the so-called
universal R-matrix [also see the graphic representation in Fig.~\ref{Fig4}]:%
\begin{align}
\exp_{q}^{\ast}(\text{i}\mathbf{p}|\hspace{0.01in}\mathbf{x})  &  =\tau
\circ\lbrack(\mathcal{R}_{[2]}^{-1}\otimes\mathcal{R}_{[1]}^{-1}%
)\triangleright\exp_{q}(\text{i}\mathbf{x}|\hspace{-0.03in}\ominus
\hspace{-0.01in}\mathbf{p})],\nonumber\\
\exp_{q}^{\ast}(\mathbf{x}|\text{i}\mathbf{p})  &  =\tau\circ\lbrack
(\mathcal{R}_{[2]}^{-1}\otimes\mathcal{R}_{[1]}^{-1})\triangleright\exp
_{q}(\ominus\hspace{0.02in}\mathbf{p}|\hspace{0.01in}\text{i}\mathbf{x})].
\label{DuaExp2}%
\end{align}
In the expressions above, $\tau$ denotes the ordinary twist operator. One can
show that the new $q$-ex\-ponentials satisfy the following eigenvalue
equations (see Fig.~\ref{Fig4}):%
\begin{align}
\exp_{q}^{\ast}(\text{i}\mathbf{p}|\hspace{0.01in}\mathbf{x})\triangleleft
\partial^{A}  &  =\text{i}p^{A}\circledast\exp_{q}^{\ast}(\text{i}%
\mathbf{p}|\hspace{0.01in}\mathbf{x}),\nonumber\\
\partial^{A}\,\bar{\triangleright}\,\exp_{q}^{\ast}(\mathbf{x}|\text{i}%
^{-1}\mathbf{p})  &  =\exp_{q}^{\ast}(\mathbf{x}|\text{i}^{-1}\mathbf{p}%
)\circledast\text{i}p^{A}. \label{EigGleExpQueAbl}%
\end{align}

Similar considerations apply to the conjugate $q$-exponentials. We only need
to modify Eqs.~(\ref{DuaExp2}) and (\ref{EigGleExpQueAbl}) by performing the
following substitutions:%
\begin{gather}
\exp_{q}^{\ast}\rightarrow\overline{\exp}_{q}^{\ast},\qquad\mathcal{R}%
_{[2]}^{-1}\otimes\mathcal{R}_{[1]}^{-1}\rightarrow\mathcal{R}_{[1]}%
\otimes\mathcal{R}_{[2]},\qquad\ominus\,\rightarrow\,\bar{\ominus},\nonumber\\
\bar{\triangleright}\,\rightarrow\,\triangleright,\qquad\triangleleft
\,\rightarrow\,\bar{\triangleleft},\qquad\partial^{A}\rightarrow\hat{\partial
}^{A}.
\end{gather}

The $q$-exponentials in Eq.~(\ref{DuaExp2}) are related to the conjugate
$q$-exponentials. To see this, we rewrite the eigenvalue equations in
(\ref{EigGleExpQueAbl}) by using the identity $\hat{\partial}^{A}%
=q^{6}\partial^{A}$ as follows:%
\begin{align}
\exp_{q}^{\ast}(\text{i}\mathbf{p}|\hspace{0.01in}\mathbf{x})\triangleleft
\hat{\partial}^{A}  &  =\text{i}q^{6}p^{A}\circledast\exp_{q}^{\ast}%
(\text{i}\mathbf{p}|\hspace{0.01in}\mathbf{x}),\nonumber\\
\hat{\partial}^{A}\,\bar{\triangleright}\,\exp_{q}^{\ast}(\mathbf{x}%
|\text{i}^{-1}\mathbf{p})  &  =\exp_{q}^{\ast}(\mathbf{x}|\text{i}%
^{-1}\mathbf{p})\circledast\text{i}q^{6}p^{A}.
\end{align}
These are the eigenvalue equations for $\overline{\exp}_{q}($%
i$^{-1}q^{6}\mathbf{p}|\mathbf{x})$ and $\overline{\exp}_{q}(\mathbf{x}%
|$i$q^{6}\mathbf{p})$, so the following identifications are valid:%
\begin{equation}
\exp_{q}^{\ast}(\text{i}\mathbf{p}|\mathbf{x})=\overline{\exp}_{q}%
(\text{i}^{-1}q^{6}\mathbf{p}|\mathbf{x}),\qquad\exp_{q}^{\ast}(\mathbf{x}%
|\hspace{0.01in}\text{i}^{-1}\mathbf{p})=\overline{\exp}_{q}(\mathbf{x}%
|\text{i}q^{6}\mathbf{p}). \label{IdeSteExpEpxKon1}%
\end{equation}
%

For the sake of completeness, we also write down how the $q$-ex\-ponentials of
$q$-de\-formed Euclidean space behave under quantum space conjugation:%
\begin{align}
\overline{\exp_{q}(\mathbf{x}|\text{i}\mathbf{p})}  &  =\exp_{q}(\text{i}%
^{-1}\mathbf{p}|\mathbf{x}), & \overline{\overline{\exp}_{q}(\mathbf{x}%
|\text{i}\mathbf{p})}  &  =\overline{\exp}_{q}(\text{i}^{-1}\mathbf{p}%
|\mathbf{x}),\nonumber\\
\overline{\exp_{q}^{\ast}(\text{i}\mathbf{p}|\mathbf{x})}  &  =\exp_{q}^{\ast
}(\mathbf{x}|\text{i}^{-1}\mathbf{p}), & \overline{\overline{\exp}_{q}^{\ast
}(\text{i}\mathbf{p}|\mathbf{x})}  &  =\overline{\exp}_{q}^{\ast}%
(\mathbf{x}|\text{i}^{-1}\mathbf{p}). \label{KonEigExpQua}%
\end{align}

\section{Hamilton operator for a free particle\label{KapHerSchGle}}

Since the $q$-de\-formed Hamilton operator of a free nonrelativistic
particle is supposed to be invariant under rotations, it must behave like a
scalar concerning the action of the Hopf algebra $\mathcal{U}_{q}%
(\operatorname*{su}\nolimits_{2})$. For this reason, we choose the following
expression as Hamilton operator for a free nonrelativistic particle with mass
$m$:%
\begin{equation}
H_{0}=-(2\hspace{0.01in}m)^{-1}g_{AB}\hspace{0.01in}\partial^{A}\partial^{B}=-(2\hspace
{0.01in}m)^{-1}\partial^{A}\partial_{A}. \label{Ham2}%
\end{equation}

Due to its definition, the Hamilton operator $H_{0}$ is a central element of
the algebra of $q$-de\-formed partial derivatives:%
\begin{equation}
\lbrack H_{0},\partial^{A}]=0,\quad A\in\{+,3,-\}. \label{ComHP}%
\end{equation}
The conjugation properties of the partial derivatives imply that $H_{0}$ is invariant under conjugation [cf.
Eq.~(\ref{KonAbl}) of Chap.~\ref{KapParDer}]:%
\begin{equation}
\overline{H_{0}}=H_{0}. \label{RelBedHamFre}%
\end{equation}

We mention that $H_{0}$ results from the low-energy limit of the following
energy-momentum relation:%
\begin{equation}
E_{\mathbf{p}}^{\hspace{0.01in}2}=c^{2}(\hspace{0.01in}p^{A}p_{A}+(m\hspace{0.01in}c)^{2}).
\end{equation}
You can see this by the following calculation:%
\begin{align}
E_{\mathbf{p}}  &  =c\hspace{0.01in}\sqrt{p^{A}p_{A}+(m\hspace{0.01in}c)^{2}}=m\hspace
{0.01in}c^{2}\sqrt{1+(m\hspace{0.01in}c)^{-2}p^{A}p_{A}}\nonumber\\
&  =m\hspace{0.01in}c^{2}(1+2^{-1}(m\hspace{0.01in}c)^{-2}p^{A}p_{A}%
+\ldots)\nonumber\\
&  =m\hspace{0.01in}c^{2}+(2\hspace{0.01in}m)^{-1}p^{A}p_{A}+\ldots
\label{LimEneImpBez}%
\end{align}
The second term of the last expression in Eq.~(\ref{LimEneImpBez}) gives
$H_{0}$ if we replace the momentum variable $p^{A}$ with the operator
i$^{-1}\partial^{A}$.

\section{Solutions to the free Schr\"{o}\-dinger equations\label{LoeSchGleKap}}

In Ref.~\cite{Wachter:2020A},\ we have derived Schr\"{o}\-dinger equations for
the three-di\-men\-sion\-al $q$-de\-formed Euclidean space $\mathbb{R}_{q}^{3}$. Now
we want to find solutions to these Schr\"{o}\-dinger equations with the free
Hamilton operator given by the expression in Eq.~(\ref{Ham2}) of the previous
chapter:%
\begin{align}
\text{i}\partial_{t}\triangleright\phi_{R}(\mathbf{x},t)  &  =H_{0}%
\triangleright\phi_{R}(\mathbf{x},t),\nonumber\\
\phi_{L}^{\ast}(\mathbf{x},t)\triangleleft\partial_{t}\text{i}  &  =\phi
_{L}^{\ast}(\mathbf{x},t)\triangleleft H_{0}. \label{FreParSch1N}%
\end{align}

Due to Eq.~(\ref{ComHP}) of the previous chapter, the free Hamilton operator
commutes with the momentum operator i$^{-1}\partial_{A}$. So we seek solutions
that are eigenfunctions of the momentum operator ($A\in\{+,3,-\}$):%
\begin{align}
\text{i}^{-1}\partial_{A}\triangleright u_{\hspace{0.01in}\mathbf{p}%
}(\mathbf{x},t)  &  =u_{\hspace{0.01in}\mathbf{p}}(\mathbf{x},t)\circledast
p_{A},\nonumber\\
(u^{\ast})_{\mathbf{p}}(\mathbf{x},t)\triangleleft\partial_{A}\text{i}^{-1}
&  =p_{A}\circledast(u^{\ast})_{\mathbf{p}}(\mathbf{x},t).
\label{ImpEigWelSol}%
\end{align}
Resulting from these identities, we can write the Schr\"{o}\-dinger equation for the wave function
$u_{\hspace{0.01in}\mathbf{p}}(\mathbf{x},t)$ or $(u^{\ast})_{\mathbf{p}%
}(\mathbf{x},t)$ as follows:\footnote{For the squared momentum holds
$\mathbf{p}^{2}=p^{A}\hspace{-0.01in}\circledast p_{A}$.}%
\begin{align}
\text{i}\partial_{t}\triangleright u_{\hspace{0.01in}\mathbf{p}}%
(\mathbf{x},t)  &  =H_{0}\triangleright u_{\hspace{0.01in}\mathbf{p}%
}(\mathbf{x},t)=-(2\hspace{0.01in}m)^{-1}\hspace{0.01in}\partial^{A}%
\partial_{A}\triangleright u_{\hspace{0.01in}\mathbf{p}}(\mathbf{x}%
,t)\nonumber\\
&  =u_{\hspace{0.01in}\mathbf{p}}(\mathbf{x},t)\circledast\mathbf{p}%
^{2}(2\hspace{0.01in}m)^{-1},\label{FreSchGlImp0}\\[0.08in]
(u^{\ast})_{\mathbf{p}}(\mathbf{x},t)\triangleleft\partial_{t}\text{i}  &
=(u^{\ast})_{\mathbf{p}}(\mathbf{x},t)\triangleleft H_{0}=-(u^{\ast
})_{\mathbf{p}}(\mathbf{x},t)\triangleleft\partial^{A}\partial_{A}(2\hspace{0.01in}%
m)^{-1}\nonumber\\
&  =(2\hspace{0.01in}m)^{-1}\hspace{0.01in}\mathbf{p}^{2}\circledast(u^{\ast
})_{\mathbf{p}}(\mathbf{x},t). \label{FreSchGlImp1}%
\end{align}
The equations above show us that $u_{\hspace{0.01in}\mathbf{p}}(\mathbf{x},t)$
and $(u^{\ast})_{\mathbf{p}}(\mathbf{x},t)$ are eigenfunctions of the energy
operator as well.

To find expressions for the functions $u_{\hspace{0.01in}\mathbf{p}%
}(\mathbf{x},t)$ and $(u^{\ast})_{\mathbf{p}}(\mathbf{x},t)$, we consider the
$q$-de\-formed momentum eigenfunctions introduced in Ref.~\cite{Wachter:2019A}%
. These momentum eigenfunctions satisfy the following eigenvalue equations:%
\begin{align}
\text{i}^{-1}\partial^{A}\triangleright u_{\hspace{0.01in}\mathbf{p}%
}(\mathbf{x})  &  =u_{\hspace{0.01in}\mathbf{p}}(\mathbf{x})\circledast p^{A},
& u^{\mathbf{p}}(\mathbf{x})\,\bar{\triangleleft}\,\partial^{A}\hspace
{0.01in}\text{i}^{-1}  &  =p^{A}\circledast u^{\mathbf{p}}(\mathbf{x}%
),\nonumber\\
\text{i}^{-1}\hat{\partial}^{A}\,\bar{\triangleright}\,\bar{u}_{\hspace
{0.01in}\mathbf{p}}(\mathbf{x})  &  =\bar{u}_{\hspace{0.01in}\mathbf{p}%
}(\mathbf{x})\circledast p^{A}, & \bar{u}^{\mathbf{p}}(\mathbf{x}%
)\triangleleft\hat{\partial}^{A}\hspace{0.01in}\text{i}^{-1}  &
=p^{A}\circledast\bar{u}^{\mathbf{p}}(\mathbf{x}).
\label{EigGleImpOpeImpEigFkt0}%
\end{align}
Since the $q$-exponentials of Chap.~\ref{KapExp} are eigenfunctions of
$q$-de\-formed partial derivatives, the $q$-de\-formed momentum eigenfunctions
can take on the following form:%
\begin{align}
u_{\hspace{0.01in}\mathbf{p}}(\mathbf{x})  &  =\operatorname*{vol}%
\nolimits^{-1/2}\exp_{q}(\mathbf{x}|\text{i}\mathbf{p}), & u^{\mathbf{p}%
}(\mathbf{x})  &  =\operatorname*{vol}\nolimits^{-1/2}\exp_{q}(\text{i}%
^{-1}\mathbf{p}|\hspace{0.01in}\mathbf{x}),\nonumber\\
\bar{u}_{\hspace{0.01in}\mathbf{p}}(\mathbf{x})  &  =\operatorname*{vol}%
\nolimits^{-1/2}\overline{\exp}_{q}(\mathbf{x}|\text{i}\mathbf{p}), & \bar
{u}^{\mathbf{p}}(\mathbf{x})  &  =\operatorname*{vol}\nolimits^{-1/2}%
\overline{\exp}_{q}(\text{i}^{-1}\mathbf{p}|\hspace{0.01in}\mathbf{x}).
\label{ImpEigFktqDef}%
\end{align}
The volume element $\operatorname*{vol}$ is defined by the expression in
Eq.~(\ref{VolEleDef}) of the next chapter. We can also introduce
dual\ momentum eigenfunctions [cf. Eq.~(\ref{DuaExp2}) of Chap.~\ref{KapExp}]:%
\begin{align}
(u^{\ast})_{\mathbf{p}}(\mathbf{x})  &  =\operatorname*{vol}\nolimits^{-1/2}%
\exp_{q}^{\ast}(\text{i}\mathbf{p}|\hspace{0.01in}\mathbf{x}), & (u^{\ast
})^{\mathbf{p}}(\mathbf{x})  &  =\operatorname*{vol}\nolimits^{-1/2}\exp
_{q}^{\ast}(\mathbf{x}|\text{i}^{-1}\mathbf{p}),\nonumber\\
(\bar{u}^{\ast})_{\mathbf{p}}(\mathbf{x})  &  =\operatorname*{vol}%
\nolimits^{-1/2}\overline{\exp}_{q}^{\ast}(\text{i}\mathbf{p}|\hspace
{0.01in}\mathbf{x}), & (\bar{u}^{\ast})^{\mathbf{p}}(\mathbf{x})  &
=\operatorname*{vol}\nolimits^{-1/2}\overline{\exp}_{q}^{\ast}(\mathbf{x}%
|\text{i}^{-1}\mathbf{p}). \label{DefDuaImpEigFktWdh}%
\end{align}
The corresponding eigenvalue equations are given by [cf.
Eq.~(\ref{EigGleExpQueAbl}) of Chap.~\ref{KapExp}]%
\begin{align}
(u^{\ast})_{\mathbf{p}}(\mathbf{x})\triangleleft\partial^{A}\hspace
{0.01in}\text{i}^{-1}\hspace{-0.01in}  &  =p^{A}\circledast(u^{\ast}%
)_{\mathbf{p}}(\mathbf{x}),\nonumber\\
\text{i}^{-1}\partial^{A}\,\bar{\triangleright}\,(u^{\ast})^{\mathbf{p}%
}(\mathbf{x})  &  =(u^{\ast})^{\mathbf{p}}(\mathbf{x})\circledast p^{A},
\label{ImpEigFktqDef2}%
\end{align}
or%
\begin{align}
(\bar{u}^{\ast})_{\mathbf{p}}(\mathbf{x})\,\bar{\triangleleft}\,\hat{\partial
}^{A}\hspace{0.01in}\text{i}^{-1}\hspace{-0.01in}  &  =p^{A}\circledast
(\bar{u}^{\ast})_{\mathbf{p}}(\mathbf{x}),\nonumber\\
\text{i}^{-1}\hat{\partial}^{A}\triangleright(\bar{u}^{\ast})^{\mathbf{p}%
}(\mathbf{x})  &  =(\bar{u}^{\ast})^{\mathbf{p}}(\mathbf{x})\circledast p^{A}.
\end{align}

In what follows, we restrict our considerations to the momentum eigenfunctions
$u_{\hspace{0.01in}\mathbf{p}}(\mathbf{x})$ and $(u^{\ast})_{\mathbf{p}%
}(\mathbf{x})$.$\ $We can obtain the results for the other momentum
eigenfunctions by simple substitutions specified at the end of this chapter.

We have shown in Ref.~\cite{Wachter:2020A} that the time evolution operator
for the quantum space $\mathbb{R}_{q}^{3}$ is of the same form as in the
undeformed case. For this reason, we get solutions to our $q$-de\-formed
Schr\"{o}\-din\-ger equations by applying the operators $\exp(-$i$tH_{0})$ and
$\exp($i$tH_{0})$ to time-independent functions $\phi_{R}(\mathbf{x},0)$ and
$\phi_{L}^{\ast}(\mathbf{x},0)$:%
\begin{align}
\phi_{R}(\mathbf{x},t)  &  =\exp(-\text{i}tH_{0})\triangleright\phi
_{R}(\mathbf{x},0),\nonumber\\
\phi_{L}^{\ast}(\mathbf{x},t)  &  =\phi_{L}^{\ast}(\mathbf{x},0)\triangleleft
\exp(\text{i}H_{0}t). \label{AnwZeitEnt}%
\end{align}
In the same way, we can obtain plane wave solutions to our Schr\"{o}\-dinger
equations from the momentum eigenfunctions $u_{\hspace{0.01in}\mathbf{p}%
}(\mathbf{x})$ and $(u^{\ast})_{\mathbf{p}}(\mathbf{x})$, i.~e.%
\begin{align}
u_{\hspace{0.01in}\mathbf{p}}(\mathbf{x},t)  &  =\exp(-\text{i}tH_{0}%
)\triangleright u_{\hspace{0.01in}\mathbf{p}}(\mathbf{x})=u_{\hspace
{0.01in}\mathbf{p}}(\mathbf{x})\circledast\exp(-\text{i}t\hspace{0.01in}%
\mathbf{p}^{2}(2\hspace{0.01in}m)^{-1})\nonumber\\
&  =\operatorname*{vol}\nolimits^{-1/2}\exp_{q}(\mathbf{x}|\text{i}%
\mathbf{p})\circledast\exp(-\text{i}t\hspace{0.01in}\mathbf{p}^{2}%
(2\hspace{0.01in}m)^{-1}), \label{ConPlaWav0}%
\end{align}
and%
\begin{align}
(u^{\ast})_{\mathbf{p}}(\mathbf{x},t)  &  =(u^{\ast})_{\mathbf{p}}%
(\mathbf{x})\triangleleft\exp(\text{i}H_{0}t)=\exp(\text{i}t\hspace
{0.01in}\mathbf{p}^{2}(2\hspace{0.01in}m)^{-1})\circledast(u^{\ast
})_{\mathbf{p}}(\mathbf{x})\nonumber\\
&  =\operatorname*{vol}\nolimits^{-1/2}\exp(\text{i}t\hspace{0.01in}%
\mathbf{p}^{2}(2\hspace{0.01in}m)^{-1})\circledast\exp_{q}^{\ast}%
(\text{i}\mathbf{p|\hspace{0.01in}x}). \label{ConPlaWav}%
\end{align}

The momentum eigenfunctions are multiplied by a time-dependent phase factor if
the time evolution operator acts on them. This phase factor is given by%
\begin{equation}
\exp(\pm\hspace{0.01in}\text{i}t\hspace{0.01in}\mathbf{p}^{2}(2\hspace{0.01in}m)^{-1}%
)=\sum_{k\hspace{0.01in}=\hspace{0.01in}0}^{\infty}\frac{1}{k!}\left(
\pm\hspace{0.01in}\text{i}t(2\hspace{0.01in}m)^{-1}\right)  ^{k}\mathbf{p}^{2k},
\label{PhaFac}%
\end{equation}
where powers of $\mathbf{p}^{2}(=g^{AB}\hspace{0.02in}p_{A}\circledast p_{B})$
are calculated by using the star-prod\-uct:%
\begin{equation}
\mathbf{p}^{2k}=\hspace{0.01in}\overset{k-\text{times}}{\overbrace
{\mathbf{p}^{2}\circledast\ldots\circledast\mathbf{p}^{2}}}\hspace
{0.01in}=\sum_{l\hspace{0.01in}=\hspace{0.01in}0}^{k}\hspace{0.01in}%
(C_{q})_{l}^{k}\,(\hspace{0.01in}p_{-})^{k-l}(\hspace{0.01in}p_{3}%
)^{2l}(\hspace{0.01in}p_{+})^{k-l}, \label{EntPotP}%
\end{equation}
The coefficients $(C_{q})_{l}^{k}$ in the series expansion above satisfy the
following recurrence relation ($\lambda_{+}=q+q^{-1}$):%
\begin{equation}
(C_{q})_{l}^{k}=-\lambda_{+}\hspace{0.02in}q^{4l}(C_{q})_{l}^{k\hspace
{0.01in}-1}\hspace{-0.01in}+q^{-2}(C_{q})_{l\hspace{0.01in}-1}^{k\hspace
{0.01in}-1}.
\end{equation}
As you can verify by inserting, this recurrence relation has the following
solution:%
\begin{equation}
(C_{q})_{l}^{k}=q^{-2l}(-\lambda_{+})^{k-l}%
\genfrac{[}{]}{0pt}{}{k}{l}%
_{q^{4}}.
\end{equation}
The $q$-de\-formed binomial coefficients are defined in complete analogy to
the undeformed case:%
\begin{equation}%
\genfrac{[}{]}{0pt}{}{n}{k}%
_{q}=\frac{[[\hspace{0.01in}n]]_{q}!}{[[\hspace{0.01in}n-k]]_{q}%
!\hspace{0.01in}[[k]]_{q}!}. \label{qBinKoeBas}%
\end{equation}
Combining our results, we finally get:%
\begin{gather}
\exp(\pm\hspace{0.01in}\text{i}t\hspace{0.01in}\mathbf{p}^{2}(2\hspace{0.01in}m)^{-1}%
)=\nonumber\\
=\sum_{k\hspace{0.01in}=\hspace{0.01in}0}^{\infty}\frac{1}{k!}\left(
\frac{\pm\hspace{0.01in}\text{i}t}{2\hspace{0.01in}m}\right)  ^{k}\sum_{l\hspace
{0.01in}=\hspace{0.01in}0}^{k}q^{-2l}(-\lambda_{+})^{k-l}%
\genfrac{[}{]}{0pt}{}{k}{l}%
_{q^{4}}(\hspace{0.01in}p_{-})^{k-l}(\hspace{0.01in}p_{3})^{2l}(\hspace
{0.01in}p_{+})^{k-l}\nonumber\\
=\sum_{k\hspace{0.01in}=\hspace{0.01in}0}^{\infty}\frac{1}{k!}\left(
\frac{\mp\hspace{0.01in}\text{i}t\lambda_{+}}{2\hspace{0.01in}m}\hspace{0.01in}p_{-}%
p_{+}\right)  ^{k}\frac{1}{((\hspace{0.01in}p_{3})^{2}/(-\hspace{0.01in}%
q^{2}\lambda_{+}\hspace{0.01in}p_{-}p_{+});q^{4})_{k}}.
\end{gather}
The second identity is a consequence of Heine's binomial formula
\cite{Kac:2002eb}:%
\begin{equation}
\frac{1}{(z\hspace{0.01in};q)_{k}}=\frac{1}{(1-z)(1-z\hspace{0.01in}%
q)\ldots(1-z\hspace{0.01in}q^{k-1})}=\sum_{l\hspace{0.01in}=\hspace{0.01in}%
0}^{k}%
\genfrac{[}{]}{0pt}{}{k}{l}%
_{q}\hspace{0.01in}z^{l}. \label{HeiBinFor}%
\end{equation}

Due to Eqs.~(\ref{ConPlaWav0}) and (\ref{ConPlaWav}), we must calculate the
star-prod\-uct of the time-dependent phase factor and the time-independent
momentum eigenfunction in the end. To get an expression for $u_{\hspace
{0.01in}\mathbf{p}}(\mathbf{x},t)$, for example, we proceed as follows:%
\begin{align}
&  \hspace{0.01in}\mathbf{p}^{2k}\circledast(\hspace{0.01in}p_{-})^{n_{-}%
}(\hspace{0.01in}p_{3})^{n_{3}}(\hspace{0.01in}p_{+})^{n_{+}}=\,(\hspace
{0.01in}p_{-})^{n_{-}}\hspace{-0.01in}\circledast\mathbf{p}^{2k}%
\circledast(\hspace{0.01in}p_{3})^{n_{3}}(\hspace{0.01in}p_{+})^{n_{+}%
}\nonumber\\
&  \qquad=\,\sum_{l\hspace{0.01in}=\hspace{0.01in}0}^{k}\hspace{0.01in}%
(C_{q})_{l}^{k}\,(\hspace{0.01in}p_{-})^{n_{-}+\hspace{0.01in}k-l}%
(\hspace{0.01in}p_{3})^{2l}(\hspace{0.01in}p_{+})^{k-l}\circledast
(\hspace{0.01in}p_{3})^{n_{3}}(\hspace{0.01in}p_{+})^{n_{+}}\nonumber\\
&  \qquad=\,\sum_{l\hspace{0.01in}=\hspace{0.01in}0}^{k}\hspace{0.01in}%
q^{2n_{3}(k-l)}(C_{q})_{l}^{k}\,(\hspace{0.01in}p_{-})^{n_{-}+\hspace
{0.01in}k-l}(\hspace{0.01in}p_{3})^{n_{3}+2l}(\hspace{0.01in}p_{+}%
)^{n_{+}+\hspace{0.01in}k-l}. \label{ZwiStePro}%
\end{align}
In the first step of the calculation above, we have used the fact that
$\mathbf{p}^{2}$ is a central element of the momentum algebra. In the second
step, we have inserted the expression given in Eq.~(\ref{EntPotP}). The last
step follows from Eq.~(\ref{StaProForExp}) in Chap.~\ref{KapQuaZeiEle} if we
take into account that $p_{A}=g_{AB}\hspace{0.01in}p^{B}$. With the result of Eq.~(\ref{ZwiStePro}), we obtain
from Eqs.~(\ref{ConPlaWav0}) and (\ref{PhaFac}) together with
Eq.~(\ref{ExpEukExp}) of Chap.~\ref{KapExp} the following expression for
$u_{\hspace{0.01in}\mathbf{p}}(\mathbf{x},t)$:%
\begin{align}
u_{\hspace{0.01in}\mathbf{p}}(\mathbf{x},t)=  &  \operatorname*{vol}%
\nolimits^{-1/2}\sum_{\underline{n}\hspace{0.01in}=\hspace{0.01in}0}^{\infty
}\sum_{k\hspace{0.01in}=\hspace{0.01in}0}^{\infty}\sum_{l\hspace
{0.01in}=\hspace{0.01in}0}^{k}\frac{(-\lambda_{+})^{k-l}q^{-2l+2n_{3}(k-l)}%
}{k!\,[[\hspace{0.01in}n_{+}]]_{q^{4}}!\,[[\hspace{0.01in}n_{3}]]_{q^{2}%
}!\,[[\hspace{0.01in}n_{-}]]_{q^{4}}!}\,%
\genfrac{[}{]}{0pt}{}{k}{l}%
_{q^{4}}\nonumber\\
&  \qquad\quad\times(2\hspace{0.01in}m)^{-k}(\text{i}t)^{k}(x^{+})^{n_{+}%
}(x^{3})^{n_{3}}(x^{-})^{n_{-}}\nonumber\\
&  \qquad\quad\times(\text{i}p_{-})^{n_{-}+\hspace{0.01in}k-l}(\text{i}%
p_{3})^{n_{3}+2l}(\text{i}p_{+})^{n_{+}+\hspace{0.01in}k-l}.
\label{ExpForEbeWel}%
\end{align}

The time-dependent phase factor depends on $\mathbf{p}^{2}$. Thus the phase
factor is a central element of the $q$-de\-formed momentum algebra. With this
insight, we can show that our plane wave solutions are momentum eigenfunctions
as well [also see Eq.~(\ref{ImpEigWelSol})]:%
\begin{align}
\text{i}^{-1}\partial_{A}\triangleright u_{\hspace{0.01in}\mathbf{p}%
}(\mathbf{x},t)  &  =\text{i}^{-1}\partial_{A}\triangleright u_{\hspace
{0.01in}\mathbf{p}}(\mathbf{x})\circledast\exp\left(  -\text{i}t\hspace
{0.01in}\mathbf{p}^{2}(2\hspace{0.01in}m)^{-1}\right) \nonumber\\
&  =u_{\hspace{0.01in}\mathbf{p}}(\mathbf{x})\circledast p_{A}\circledast
\exp\left(  -\text{i}t\hspace{0.01in}\mathbf{p}^{2}(2\hspace{0.01in}%
m)^{-1}\right) \nonumber\\
&  =u_{\hspace{0.01in}\mathbf{p}}(\mathbf{x})\circledast\exp\left(
-\text{i}t\hspace{0.01in}\mathbf{p}^{2}(2\hspace{0.01in}m)^{-1}\right)
\circledast p_{A}=u_{\hspace{0.01in}\mathbf{p}}(\mathbf{x},t)\circledast
p_{A}.
\end{align}

By quantum space conjugation, you can obtain further $q$-de\-formed
Schr\"{o}\-dinger equations from Eq.~(\ref{FreParSch1N}), i.~e. [also see
Eq.~(\ref{RegConAbl}) of Chap.~\ref{KapParDer}]%
\begin{align}
\phi_{L}(\mathbf{x},t)\,\bar{\triangleleft}\,\partial_{t}\text{i}  &
=\phi_{L}(\mathbf{x},t)\,\bar{\triangleleft}\,H_{0},\nonumber\\
\text{i}\partial_{t}\,\bar{\triangleright}\,\phi_{R}^{\ast}(\mathbf{x},t)  &
=H_{0}\,\bar{\triangleright}\,\phi_{R}^{\ast}(\mathbf{x},t) \label{KonSchrGle}%
\end{align}
with%
\begin{equation}
\overline{\phi_{R}(\mathbf{x},t)}=\phi_{L}(\mathbf{x},t),\qquad\overline
{\phi_{L}^{\ast}(\mathbf{x},t)}=\phi_{R}^{\ast}(\mathbf{x},t).
\label{VerKonWelFkt}%
\end{equation}
Accordingly, the quantum space conjugates of the plane waves $u_{\hspace
{0.01in}\mathbf{p}}(\mathbf{x},t)$ and $(u^{\ast})_{\mathbf{p}}(\mathbf{x},t)$
are plane wave solutions to the $q$-de\-formed Schr\"{o}\-dinger equations given
in Eq.~(\ref{KonSchrGle}), i.~e.%
\begin{align}
u^{\mathbf{p}}(\mathbf{x},t)\,\bar{\triangleleft}\,\partial_{t}\text{i}  &
=u^{\mathbf{p}}(\mathbf{x},t)\,\bar{\triangleleft}\,H_{0}=(2\hspace
{0.01in}m)^{-1}\mathbf{p}^{2}\circledast u^{\mathbf{p}}(\mathbf{x}%
,t),\nonumber\\
\text{i}\partial_{t}\,\bar{\triangleright}\,(u^{\ast})^{\mathbf{p}}%
(\mathbf{x},t)  &  =H_{0}\,\bar{\triangleright}\,(u^{\ast})^{\mathbf{p}%
}(\mathbf{x},t)=(u^{\ast})^{\mathbf{p}}(\mathbf{x},t)\circledast\mathbf{p}%
^{2}(2\hspace{0.01in}m)^{-1}%
\end{align}
with%
\begin{equation}
u^{\mathbf{p}}(\mathbf{x},t)=\overline{u_{\hspace{0.01in}\mathbf{p}%
}(\mathbf{x},t)},\qquad(u^{\ast})^{\mathbf{p}}(\mathbf{x},t)=\overline
{(u^{\ast})_{\mathbf{p}}(\mathbf{x},t)}. \label{KonEbeWel}%
\end{equation}
The new plane wave solutions are subject to the identities%
\begin{align}
u^{\mathbf{p}}(\mathbf{x},t)  &  =u^{\mathbf{p}}(\mathbf{x})\,\bar
{\triangleleft}\,\exp(\text{i}H_{0}t)\nonumber\\
&  =\exp\left(  \text{i}t\hspace{0.01in}\mathbf{p}^{2}(2\hspace{0.01in}%
m)^{-1}\right)  \circledast u^{\mathbf{p}}(\mathbf{x})\nonumber\\
&  =\exp\left(  \text{i}t\hspace{0.01in}\mathbf{p}^{2}(2\hspace{0.01in}%
m)^{-1}\right)  \circledast\exp_{q}(\text{i}^{-1}\mathbf{p}|\hspace
{0.01in}\mathbf{x})\operatorname*{vol}\nolimits^{-1/2}%
\end{align}
and%
\begin{align}
(u^{\ast})^{\mathbf{p}}(\mathbf{x},t)  &  =\exp(-\text{i}tH_{0})\,\bar
{\triangleright}\,(u^{\ast})^{\mathbf{p}}(\mathbf{x})\nonumber\\
&  =(u^{\ast})^{\mathbf{p}}(\mathbf{x})\circledast\exp\left(  -\text{i}%
t\hspace{0.01in}\mathbf{p}^{2}(2\hspace{0.01in}m)^{-1}\right) \nonumber\\
&  =\operatorname*{vol}\nolimits^{-1/2}\exp_{q}^{\ast}(\mathbf{x}%
|\text{i}^{-1}\mathbf{p})\circledast\exp\left(  -\text{i}t\hspace
{0.01in}\mathbf{p}^{2}(2\hspace{0.01in}m)^{-1}\right)  .
\end{align}
Last not but least, we write down an explicit formula for $u^{\mathbf{p}%
}(\mathbf{x},t)$:%
\begin{align}
u^{\mathbf{p}}(\mathbf{x},t)=  &  \operatorname*{vol}\nolimits^{-1/2}%
\sum_{\underline{n}\hspace{0.01in}=\hspace{0.01in}0}^{\infty}\sum
_{k\hspace{0.01in}=\hspace{0.01in}0}^{\infty}\sum_{l\hspace{0.01in}%
=\hspace{0.01in}0}^{k}\frac{(-\lambda_{+})^{k-l}q^{-2l+2n_{3}(k-l)}%
}{k!\,[[\hspace{0.01in}n_{+}]]_{q^{4}}!\,[[\hspace{0.01in}n_{3}]]_{q^{2}%
}!\,[[\hspace{0.01in}n_{-}]]_{q^{4}}!}\,%
\genfrac{[}{]}{0pt}{}{k}{l}%
_{q^{4}}\nonumber\\
&  \qquad\quad\times(2\hspace{0.01in}m)^{-k}(\text{i}^{-1}p_{-})^{n_{-}%
+\hspace{0.01in}k-l}(\text{i}^{-1}p_{3})^{n_{3}+2l}(\text{i}^{-1}p_{+}%
)^{n_{+}+\hspace{0.01in}k-l}\nonumber\\
&  \qquad\quad\otimes(\text{i}^{-1}t)^{k}(x^{+})^{n_{+}}(x^{3})^{n_{3}}%
(x^{-})^{n_{-}}.
\end{align}

Once again, the plane wave solutions $u^{\mathbf{p}}(\mathbf{x},t)$ and
$(u^{\ast})^{\mathbf{p}}(\mathbf{x},t)$ describe free particle states with
definite energy and momentum. Due to Eqs.~(\ref{EigGleImpOpeImpEigFkt0}) and
(\ref{ImpEigFktqDef2}), it holds%
\begin{align}
u^{\mathbf{p}}(\mathbf{x},t)\,\bar{\triangleleft}\,\partial^{A}\text{i}^{-1}
&  =\hspace{0.01in}p^{A}\circledast u^{\mathbf{p}}(\mathbf{x},t),\nonumber\\
u^{\mathbf{p}}(\mathbf{x},t)\,\bar{\triangleleft}\,H_{0}  &  =-\hspace
{0.01in}u^{\mathbf{p}}(\mathbf{x},t)\,\bar{\triangleleft}\,\partial
^{A}\partial_{A}(2\hspace{0.01in}m)^{-1}\nonumber\\
&  =(2\hspace{0.01in}m)^{-1}\hspace{0.01in}\mathbf{p}^{2}\circledast
u^{\mathbf{p}}(\mathbf{x},t)
\end{align}
and%
\begin{align}
\text{i}^{-1}\partial^{A}\,\bar{\triangleright}\,(u^{\ast})^{\mathbf{p}%
}(\mathbf{x},t)  &  =(u^{\ast})^{\mathbf{p}}(\mathbf{x},t)\circledast
p^{A},\nonumber\\
H_{0}\,\bar{\triangleright}\,(u^{\ast})^{\mathbf{p}}(\mathbf{x},t)  &
=-(2\hspace{0.01in}m)^{-1}\hspace{0.01in}\partial^{A}\partial_{A}%
\,\bar{\triangleright}\,(u^{\ast})^{\mathbf{p}}(\mathbf{x},t)\nonumber\\
&  =(u^{\ast})^{\mathbf{p}}(\mathbf{x},t)\circledast\mathbf{p}^{2}%
(2\hspace{0.01in}m)^{-1}.
\end{align}

For the sake of completeness, we provide another method to obtain
$q$-de\-formed Schr\"{o}\-dinger equations and their plane wave solutions. We
only need to apply the following substitutions to the identities of the
present chapter:%
\begin{gather}
\triangleright\,\leftrightarrow\,\bar{\triangleright},\qquad\triangleleft
\,\leftrightarrow\,\bar{\triangleleft},\qquad\partial^{A}\,\leftrightarrow
\,\hat{\partial}^{A},\qquad u\,\leftrightarrow\,\bar{u},\nonumber\\
+\,\leftrightarrow\,-,\qquad q\,\leftrightarrow\,q^{-1}.
\end{gather}
Due to these substitutions, we will not consider the momentum eigenfunctions
$\bar{u}_{\hspace{0.01in}\mathbf{p}}$ and $(\bar{u}^{\ast})_{\hspace
{0.01in}\mathbf{p}}$ or $\bar{u}^{\mathbf{p}}$ and $(\bar{u}^{\ast
})^{\mathbf{p}}$ in the following.

\section{Orthonormality and completeness\label{KapOrtVolEBeWel}}

The $q$-de\-formed momentum eigenfunctions [cf. Eqs.~(\ref{ImpEigFktqDef}) and
(\ref{DefDuaImpEigFktWdh}) of the previous chapter] form a complete
orthonormal system of functions \cite{Kempf:1994yd,Wachter:2019A}. In the
following, we will show that the same applies to the $q$-de\-formed plane
waves derived in the previous chapter as solutions to the free Schr\"{o}\-dinger equations.

We recall that the $q$-de\-formed momentum eigenfunctions fulfill the
orthogonality relation \cite{Wachter:2019A}%
\begin{align}
\int\text{d}_{q}^{3}x\,(u^{\ast})_{\mathbf{p}}(\mathbf{x})\circledast
u_{\hspace{0.01in}\mathbf{p}^{\prime}}(\mathbf{x})  &  =\operatorname*{vol}%
\nolimits^{-1}\hspace{-0.02in}\int\text{d}_{q}^{3}x\hspace{0.01in}\exp
_{q}^{\ast}(\text{i}\mathbf{p}|\mathbf{x})\circledast\exp_{q}(\mathbf{x}%
|\text{i}\mathbf{p}^{\prime})\nonumber\\
&  =\operatorname*{vol}\nolimits^{-1}\hspace{-0.01in}\delta_{q}^{\hspace
{0.01in}3}((\ominus\hspace{0.01in}\kappa^{-1}\mathbf{p})\oplus\mathbf{p}%
^{\prime}) \label{SkaProEbeDreExpWie0}%
\end{align}
or%
\begin{align}
\int\text{d}_{q}^{3}x\,u^{\mathbf{p}}(\mathbf{x})\circledast(u^{\ast
})^{\mathbf{p}^{\prime}}\hspace{-0.02in}(\mathbf{x})  &  =\operatorname*{vol}\nolimits^{-1}%
\hspace{-0.02in}\int\text{d}_{q}^{3}x\hspace{0.01in}\exp_{q}(\text{i}%
^{-1}\mathbf{p}|\mathbf{x})\circledast\exp_{q}^{\ast}(\mathbf{x|}\text{i}%
^{-1}\mathbf{p}^{\prime})\nonumber\\
&  =\operatorname*{vol}\nolimits^{-1}\hspace{-0.01in}\delta_{q}^{\hspace
{0.01in}3}(\hspace{0.01in}\mathbf{p}\oplus(\ominus\hspace{0.01in}\kappa
^{-1}\mathbf{p}^{\prime})).
\end{align}
We use the convention that an integral without limits is an integral over all space.
$\delta_{q}^{\hspace{0.01in}3}(\hspace{0.01in}\mathbf{p})$ denotes a
$q$-de\-formed version of the three-di\-men\-sion\-al delta function. Accordingly,
we have%
\begin{equation}
\delta_{q}^{\hspace{0.01in}3}(\hspace{0.01in}\mathbf{p})=\int\text{d}_{q}%
^{3}x\hspace{0.01in}\exp_{q}(\text{i}^{-1}\mathbf{p}|\mathbf{x})=\int
\text{d}_{q}^{3}x\hspace{0.01in}\exp_{q}^{\ast}(\mathbf{x|}\text{i}%
^{-1}\mathbf{p})
\end{equation}
and%
\begin{equation}
\operatorname*{vol}=\int\text{d}_{q}^{3}p\hspace{0.02in}\delta_{q}%
^{\hspace{0.01in}3}(\mathbf{p})=\int\text{d}_{q}^{3}p\int\text{d}_{q}%
^{3}x\hspace{0.01in}\exp_{q}(\text{i}^{-1}\mathbf{p}|\mathbf{x}).
\label{VolEleDef}%
\end{equation}
In analogy to their undeformed counterparts, the $q$-de\-formed delta
functions fulfill the following identities:\footnote{The occurrence of
$\kappa^{-1}=q^{-6}$ indicates that the spatial coordinates are multiplied by
that constant.}%
\begin{align}
f(\hspace{0.01in}\mathbf{y})  &  =\operatorname*{vol}\nolimits^{-1}%
\int\text{d}_{q}^{3}x\,\delta_{q}^{\hspace
{0.01in}3}(\hspace{0.01in}\mathbf{y}\oplus(\ominus\hspace{0.01in}\kappa
^{-1}\mathbf{x}))\circledast f(\mathbf{x})\nonumber\\
&  =\operatorname*{vol}\nolimits^{-1}\int%
\text{d}_{q}^{3}x\,\delta_{q}^{\hspace{0.01in}3}((\ominus\hspace{0.01in}%
\kappa^{-1}\mathbf{y})\oplus\mathbf{x})\circledast f(\mathbf{x})\nonumber\\
&  =\operatorname*{vol}\nolimits^{-1}\int%
\text{d}_{q}^{3}x\,f(\mathbf{x})\circledast\delta_{q}^{\hspace{0.01in}%
3}((\ominus\hspace{0.01in}\kappa^{-1}\mathbf{x})\oplus\mathbf{y}))\nonumber\\
&  =\operatorname*{vol}\nolimits^{-1}\int%
\text{d}_{q}^{3}x\,f(\mathbf{x})\circledast\delta_{q}^{\hspace{0.01in}%
3}(\mathbf{x}\oplus(\ominus\hspace{0.01in}\kappa^{-1}\mathbf{y})).
\label{AlgChaIdeqDelFkt}%
\end{align}
From Eq.~(\ref{SkaProEbeDreExpWie0}) follows that the time-dependent
$q$-de\-formed plane waves fulfill an orthonormality relation as well:%
\begin{align}
&  \int\text{d}_{q}^{3}x\,(u^{\ast})_{\mathbf{p}}(\mathbf{x},t)\circledast
u_{\hspace{0.01in}\mathbf{p}^{\prime}}(\mathbf{x},t)=\nonumber\\
&  \qquad=\int\text{d}_{q}^{3}x\hspace{0.01in}\exp(\text{i}t\hspace
{0.01in}\mathbf{p}^{2}(2\hspace{0.01in}m)^{-1})\circledast(u^{\ast
})_{\mathbf{p}}(\mathbf{x})\circledast u_{\hspace{0.01in}\mathbf{p}^{\prime}%
}(\mathbf{x})\circledast\exp(-\text{i}t\hspace{0.01in}\mathbf{p}^{\prime
2}(2\hspace{0.01in}m)^{-1})\nonumber\\
&  \qquad=\operatorname*{vol}\nolimits^{-1}\hspace{-0.01in}\exp(\text{i}%
t\hspace{0.01in}\mathbf{p}^{2}(2\hspace{0.01in}m)^{-1})\circledast\delta
_{q}^{\hspace{0.01in}3}((\ominus\hspace{0.01in}\kappa^{-1}\mathbf{p})\oplus\mathbf{p}%
^{\prime})\circledast\exp(-\text{i}t\hspace{0.01in}\mathbf{p}^{\prime
2}(2\hspace{0.01in}m)^{-1})\nonumber\\
&  \qquad=\operatorname*{vol}\nolimits^{-1}\hspace{-0.01in}\exp(\text{i}%
t\hspace{0.01in}\mathbf{p}^{2}(2\hspace{0.01in}m)^{-1})\circledast
\exp(-\text{i}t\hspace{0.01in}\mathbf{p}^{2}(2\hspace{0.01in}m)^{-1}%
)\circledast\delta_{q}^{\hspace{0.01in}3}((\ominus\hspace{0.01in}\kappa^{-1}\mathbf{p}%
)\oplus\mathbf{p}^{\prime})\nonumber\\
&  \qquad=\operatorname*{vol}\nolimits^{-1}\hspace{-0.01in}\delta_{q}%
^{\hspace{0.01in}3}((\ominus\hspace{0.01in}\kappa^{-1}\mathbf{p}%
)\oplus\mathbf{p}^{\prime}). \label{OrtRelEbeWel0Schr}
\end{align}
Likewise, it holds:%
\begin{align}
\int\text{d}_{q}^{3}x\,u^{\mathbf{p}}(\mathbf{x},t)\circledast(u^{\ast
})^{\mathbf{p}^{\prime}}\hspace{-0.02in}(\mathbf{x},t)  &  =\int\text{d}_{q}^{3}%
x\,u^{\mathbf{p}}(\mathbf{x})\circledast(u^{\ast})^{\mathbf{p}^{\prime}%
}\hspace{-0.02in}(\mathbf{x})\nonumber\\
&  =\operatorname*{vol}\nolimits^{-1}\hspace{-0.01in}\delta_{q}^{\hspace
{0.01in}3}(\hspace{0.01in}\mathbf{p}\oplus(\ominus\hspace{0.01in}\kappa
^{-1}\mathbf{p}^{\prime})). \label{OrtRelEbeWel1Schr}%
\end{align}

Let $\phi_{R}(\mathbf{x},t)$ be a solution to a $q$-de\-formed Schr\"{o}\-dinger
equation [cf. Eq.~(\ref{FreParSch1N}) of the previous chapter]. Remember that
the $q$-de\-formed momentum eigenfunctions $u_{\hspace{0.01in}\mathbf{p}%
}(\mathbf{x})$ form a complete set of functions \cite{Wachter:2019A}. Thus, we
can write the function $\phi_{R}(\mathbf{x},t=0)$ as a series expansion in terms of
these momentum eigenfunctions, i.~e.%
\begin{equation}
\phi_{R}(\mathbf{x},0)=\phi_{R}(\mathbf{x},0)=\int\text{d}_{q}^{3}%
p\,u_{\hspace{0.01in}\mathbf{p}}(\mathbf{x})\circledast c_{\hspace
{0.01in}\mathbf{p}}%
\end{equation}
with%
\begin{equation}
c_{\hspace{0.01in}\mathbf{p}}=\int\text{d}_{q}^{3}x\,(u^{\ast})_{\mathbf{p}%
}(\mathbf{x})\circledast\phi_{R}(\mathbf{x},0).
\end{equation}
For this reason, there is also a series expansion of $\phi_{R}(\mathbf{x},t)$ in
terms of the time-dependent plane waves $u_{\hspace{0.01in}\mathbf{p}%
}(\mathbf{x},t)$:%
\begin{align}
\phi_{R}(\mathbf{x},t)  &  =\exp(-\text{i}tH_{0})\triangleright\phi
_{R}(\mathbf{x},0)=\exp(-\text{i}tH_{0})\triangleright\int\text{d}_{q}%
^{3}p\,u_{\hspace{0.01in}\mathbf{p}}(\mathbf{x})\circledast c_{\hspace
{0.01in}\mathbf{p}}\nonumber\\
&  =\int\text{d}_{q}^{3}p\,\exp(-\text{i}tH_{0})\triangleright u_{\hspace
{0.01in}\mathbf{p}}(\mathbf{x})\circledast c_{\hspace{0.01in}\mathbf{p}}%
=\int\text{d}_{q}^{3}p\,u_{\hspace{0.01in}\mathbf{p}}(\mathbf{x},t)\circledast
c_{\hspace{0.01in}\mathbf{p}}. \label{ExpQR}%
\end{align}
Moreover, we can calculate the coefficients $c_{\hspace
{0.01in}\mathbf{p}}$ as follows:%
\begin{align}
\int\text{d}_{q}^{3}x\,(u^{\ast})_{\mathbf{p}}(\mathbf{x},t)\circledast
\phi_{R}(\mathbf{x},t)  &  =\int\text{d}_{q}^{3}x\,(u^{\ast})_{\mathbf{p}%
}(\mathbf{x},t)\circledast\hspace{-0.02in}\int\text{d}_{q}^{3}p^{\prime
}\,u_{\hspace{0.01in}\mathbf{p}^{\prime}}(\mathbf{x},t)\circledast
c_{\hspace{0.01in}\mathbf{p}^{\prime}}\nonumber\\
&  =\int\text{d}_{q}^{3}p^{\prime}\hspace{-0.02in}\int\text{d}_{q}%
^{3}x\,(u^{\ast})_{\mathbf{p}}(\mathbf{x},t)\circledast u_{\hspace
{0.01in}\mathbf{p}^{\prime}}(\mathbf{x},t)\circledast c_{\hspace
{0.01in}\mathbf{p}^{\prime}}\nonumber\\
&  =\int\text{d}_{q}^{3}p^{\prime}\operatorname*{vol}\nolimits^{-1}\hspace
{-0.01in}\delta_{q}^{\hspace{0.01in}3}(\hspace{0.01in}\mathbf{p}\oplus
(\ominus\hspace{0.01in}\kappa^{-1}\mathbf{p}^{\prime}))\circledast
c_{\hspace{0.01in}\mathbf{p}^{\prime}}\nonumber\\
&  =c_{\hspace{0.01in}\mathbf{p}}. \label{ExpKoeQR}%
\end{align}

The same considerations apply to the solutions of the other $q$-de\-formed versions of
the Schr\"{o}\-dinger equation. This way, we get%
\begin{equation}
\phi_{L}(\mathbf{x},t)=\int\text{d}_{q}^{3}p\,c^{\hspace{0.01in}\mathbf{p}%
}\hspace{-0.01in}\circledast u^{\mathbf{p}}(\mathbf{x},t)
\label{EntWelFktEbeDreDim1}%
\end{equation}
and%
\begin{align}
\phi_{R}^{\ast}(\mathbf{x},t)  &  =\int\text{d}_{q}^{3}p\,(u^{\ast
})^{\mathbf{p}}(\mathbf{x},t)\circledast(c^{\ast})^{\mathbf{p}}%
,\nonumber\\[0.1in]
\phi_{L}^{\ast}(\mathbf{x},t)  &  =\int\text{d}_{q}^{3}p\,(c^{\ast
})_{\mathbf{p}}\circledast(u^{\ast})_{\mathbf{p}}(\mathbf{x},t).
\label{EntWelFktEbeDreDim2}%
\end{align}
For the coefficients in the above series expansions, we have%
\begin{equation}
c^{\hspace{0.01in}\mathbf{p}}=\int\text{d}_{q}^{3}x\,\phi_{L}(\mathbf{x}%
,t)\circledast(u^{\ast})^{\mathbf{p}}(\mathbf{x},t) \label{BesEntKoeSchrEbe1}%
\end{equation}
and%
\begin{align}
(c^{\ast})_{\mathbf{p}}  &  =\int\text{d}_{q}^{3}x\,\phi_{L}^{\ast}%
(\mathbf{x},t)\circledast u_{\hspace{0.01in}\mathbf{p}}(\mathbf{x}%
,t),\nonumber\\
(c^{\ast})^{\mathbf{p}}  &  =\int\text{d}_{q}^{3}x\,u^{\mathbf{p}}%
(\mathbf{x},t)\circledast\phi_{R}^{\ast}(\mathbf{x},t).
\label{BesEntKoeSchrEbe2}%
\end{align}
The above expressions for the coefficients and the behavior of the free
Schr\"{o}\-din\-ger wave functions under quantum space conjugation [cf.
Eqs.~(\ref{VerKonWelFkt}) and (\ref{KonEbeWel}) in Chap.~\ref{LoeSchGleKap}]
imply the following conjugation properties:%
\begin{equation}
\overline{c^{\hspace{0.01in}\mathbf{p}}}=c_{\hspace{0.01in}\mathbf{p}%
},\text{\qquad}\overline{(c^{\ast})^{\mathbf{p}}}=(c^{\ast})_{\mathbf{p}}.
\label{KonBedEntKoe}%
\end{equation}

Finally, we determine \textit{completeness relations }for our $q$-de\-formed
plane waves. To this end, we consider the series expansion of $\phi_{R}%
(\mathbf{x},t)$ in terms of the plane waves $u_{\hspace{0.01in}\mathbf{p}%
}(\mathbf{x},t)$ [cf. Eq.~(\ref{ExpQR})] and insert the expression for the
 coefficients $c_{\hspace{0.01in}\mathbf{p}}$ [cf.
Eq.~(\ref{ExpKoeQR})]:%
\begin{align}
\phi_{R}(\mathbf{x},t)  &  =\int\text{d}_{q}^{3}p\,u_{\hspace{0.01in}%
\mathbf{p}}(\mathbf{x},t)\circledast c_{\hspace{0.01in}\mathbf{p}}\nonumber\\
&  =\int\text{d}_{q}^{3}p\,u_{\hspace{0.01in}\mathbf{p}}(\mathbf{x}%
,t)\circledast\hspace{-0.01in}\int\text{d}_{q}^{3}y\,(u^{\ast})_{\mathbf{p}%
}(\hspace{0.01in}\mathbf{y},t)\circledast\phi_{R}(\hspace{0.01in}%
\mathbf{y},t)\nonumber\\
&  =\int\text{d}_{q}^{3}y\int\text{d}_{q}^{3}p\,u_{\hspace{0.01in}\mathbf{p}%
}(\mathbf{x},t)\circledast(u^{\ast})_{\mathbf{p}}(\hspace{0.01in}%
\mathbf{y},t)\circledast\phi_{R}(\hspace{0.01in}\mathbf{y},t).
\label{RecVolRelZeiEbeWel1}%
\end{align}
Comparing the above result with the identities in Eq.~(\ref{AlgChaIdeqDelFkt}%
), we find the following completeness relation:%
\begin{equation}
\int\text{d}_{q}^{3}p\,u_{\hspace{0.01in}\mathbf{p}}(\mathbf{x},t)\circledast
(u^{\ast})_{\mathbf{p}}(\hspace{0.01in}\mathbf{y},t)=\operatorname*{vol}%
\nolimits^{-1}\hspace{-0.01in}\delta_{q}^{\hspace{0.01in}3}(\mathbf{x}\oplus(\ominus
\hspace{0.01in}\kappa^{-1}\mathbf{y})). \label{VolRelZeiWelDreDim1}%
\end{equation}
In the same manner, we get:%
\begin{equation}
\int\text{d}_{q}^{3}p\,(u^{\ast})^{\mathbf{p}}(\hspace{0.01in}\mathbf{y}%
,t)\circledast u^{\mathbf{p}}(\mathbf{x},t)=\operatorname*{vol}\nolimits^{-1}%
\hspace{-0.01in}\delta_{q}^{\hspace{0.01in}3}((\ominus\hspace{0.01in}\kappa^{-1}%
\mathbf{y})\oplus\mathbf{x}). \label{VolRelZeiWelDreDim2}%
\end{equation}

\section{Free particle propagators\label{KapProSchrFel}}

If we know the wave function of a quantum system at a given time, we can find
the wave function at any time with the help of the time evolution operator
[also see Eq.~(\ref{AnwZeitEnt}) of Chap.~\ref{LoeSchGleKap}]. We can also use
the propagator to solve the time evolution problem. In this chapter, we give $q$-de\-formed
expressions for the propagator of a free nonrelativistic particle.
Additionally, we are going to derive some important properties of these $q$-de\-formed propagators.

As shown in the previous chapter, we can write solutions to the $q$-de\-formed
Schr\"{o}\-din\-ger equations of a free nonrelativistic particle as a series expansion
in terms of plane waves [cf. Eqs.~(\ref{ExpQR}), (\ref{EntWelFktEbeDreDim1}),
and (\ref{EntWelFktEbeDreDim2}) of the previous chapter], i.~e.%
\begin{align}
\phi_{R}(\mathbf{x},t)  &  =\int\text{d}_{q}^{3}p\,u_{\hspace{0.01in}%
\mathbf{p}}(\mathbf{x},t)\circledast c_{\hspace{0.01in}\mathbf{p}},\nonumber\\
\phi_{L}(\mathbf{x},t)  &  =\int\text{d}_{q}^{3}p\,c^{\hspace{0.01in}%
\mathbf{p}}\hspace{-0.01in}\circledast u^{\mathbf{p}}(\mathbf{x},t),
\label{EntWicEbeWelDreDim1}%
\end{align}
and%
\begin{align}
\phi_{R}^{\ast}(\mathbf{x},t)  &  =\int\text{d}_{q}^{3}p\,(u^{\ast
})^{\mathbf{p}}(\mathbf{x},t)\circledast(c^{\ast})^{\mathbf{p}}%
,\nonumber\\[0.1in]
\phi_{L}^{\ast}(\mathbf{x},t)  &  =\int\text{d}_{q}^{3}p\,(c^{\ast
})_{\mathbf{p}}\circledast(u^{\ast})_{\mathbf{p}}(\mathbf{x},t).
\label{EntWicEbeWelDreDim2}%
\end{align}
Furthermore, we know how to calculate the corresponding coefficients
from the wave functions [cf. Eqs.~(\ref{ExpKoeQR}), (\ref{BesEntKoeSchrEbe1}),
and (\ref{BesEntKoeSchrEbe2}) of the previous chapter], i.~e.%
\begin{align}
c_{\hspace{0.01in}\mathbf{p}}  &  =\int\text{d}_{q}^{3}x\hspace{0.01in}%
(u^{\ast})_{\mathbf{p}}(\mathbf{x},t)\circledast\phi_{R}(\mathbf{x}%
,t),\nonumber\\
c^{\hspace{0.01in}\mathbf{p}}  &  =\int\text{d}_{q}^{3}x\,\phi_{L}%
(\mathbf{x},t)\circledast(u^{\ast})^{\mathbf{p}}(\mathbf{x},t),
\label{EntKoeDreDim1Wie}%
\end{align}
and%
\begin{align}
(c^{\ast})_{\mathbf{p}}  &  =\int\text{d}_{q}^{3}x\,\phi_{L}^{\ast}%
(\mathbf{x},t)\circledast u_{\hspace{0.01in}\mathbf{p}}(\mathbf{x}%
,t),\nonumber\\
(c^{\ast})^{\mathbf{p}}  &  =\int\text{d}_{q}^{3}x\,u^{\mathbf{p}}%
(\mathbf{x},t)\circledast\phi_{R}^{\ast}(\mathbf{x},t).
\label{EntKoeDreDim4Wie}%
\end{align}

Next, we derive formulas for the $q$-de\-formed propagators of the free nonrelativistic
particle. We insert the expressions from Eq.~(\ref{EntKoeDreDim1Wie}) or
Eq.~(\ref{EntKoeDreDim4Wie}) into Eq.~(\ref{EntWicEbeWelDreDim1}) or
Eq.~(\ref{EntWicEbeWelDreDim2}) and obtain the integral equations%
\begin{align}
\phi_{R}(\mathbf{x}^{\hspace{0.01in}\prime}\hspace{-0.01in},t^{\hspace
{0.01in}\prime})  &  =\int\text{d}_{q}^{3}x\,K_{R}(\mathbf{x}^{\hspace
{0.01in}\prime}\hspace{-0.01in},t^{\hspace{0.01in}\prime}\hspace
{-0.01in};\mathbf{x},t)\circledast\phi_{R}(\mathbf{x},t),\nonumber\\
\phi_{L}(\mathbf{x}^{\hspace{0.01in}\prime}\hspace{-0.01in},t^{\hspace
{0.01in}\prime})  &  =\int\text{d}_{q}^{3}x\,\phi_{L}(\mathbf{x},t)\circledast
K_{L}(\mathbf{x},t;\mathbf{x}^{\hspace{0.01in}\prime}\hspace{-0.01in}%
,t^{\hspace{0.01in}\prime}), \label{DefProDreDim1}%
\end{align}
or%
\begin{align}
\phi_{R}^{\ast}(\mathbf{x}^{\hspace{0.01in}\prime}\hspace{-0.01in}%
,t^{\hspace{0.01in}\prime})  &  =\int\text{d}_{q}^{3}x\,K_{R}^{\ast
}(\mathbf{x}^{\hspace{0.01in}\prime}\hspace{-0.01in},t^{\hspace{0.01in}\prime
}\hspace{-0.01in};\mathbf{x},t)\circledast\phi_{R}^{\ast}(\mathbf{x}%
,t),\nonumber\\[0.1in]
\phi_{L}^{\ast}(\mathbf{x}^{\hspace{0.01in}\prime}\hspace{-0.01in}%
,t^{\hspace{0.01in}\prime})  &  =\int\text{d}_{q}^{3}x\,\phi_{L}^{\ast
}(\mathbf{x},t)\circledast K_{L}^{\ast}(\mathbf{x},t;\mathbf{x}^{\hspace
{0.01in}\prime}\hspace{-0.01in},t^{\hspace{0.01in}\prime})
\label{DefProDreDim2}%
\end{align}
with the integral kernels%
\begin{align}
K_{R}(\mathbf{x}^{\hspace{0.01in}\prime}\hspace{-0.01in},t^{\hspace
{0.01in}\prime}\hspace{-0.01in};\mathbf{x},t)  &  =\int\text{d}_{q}%
^{3}p\,u_{\hspace{0.01in}\mathbf{p}}(\mathbf{x}^{\hspace{0.01in}\prime
},t^{\hspace{0.01in}\prime})\circledast(u^{\ast})_{\mathbf{p}}(\mathbf{x}%
,t),\nonumber\\
K_{L}(\mathbf{x},t;\mathbf{x}^{\hspace{0.01in}\prime}\hspace{-0.01in}%
,t^{\hspace{0.01in}\prime})  &  =\int\text{d}_{q}^{3}p\,(u^{\ast}%
)^{\mathbf{p}}(\mathbf{x},t)\circledast u^{\mathbf{p}}(\mathbf{x}%
^{\hspace{0.01in}\prime},t^{\hspace{0.01in}\prime}), \label{IntKer1}%
\end{align}
or%
\begin{align}
K_{R}^{\ast}(\mathbf{x}^{\hspace{0.01in}\prime}\hspace{-0.01in},t^{\hspace
{0.01in}\prime}\hspace{-0.01in};\mathbf{x},t)  &  =\int\text{d}_{q}%
^{3}p\,(u^{\ast})^{\mathbf{p}}(\mathbf{x}^{\hspace{0.01in}\prime}%
,t^{\hspace{0.01in}\prime})\circledast u^{\mathbf{p}}(\mathbf{x}%
,t),\nonumber\\
K_{L}^{\ast}(\mathbf{x},t;\mathbf{x}^{\hspace{0.01in}\prime}\hspace
{-0.01in},t^{\hspace{0.01in}\prime})  &  =\int\text{d}_{q}^{3}p\,u_{\hspace
{0.01in}\mathbf{p}}(\mathbf{x},t)\circledast(u^{\ast})_{\mathbf{p}}%
(\mathbf{x}^{\hspace{0.01in}\prime},t^{\hspace{0.01in}\prime}).
\label{IntKer2}%
\end{align}
Comparing Eq.~(\ref{IntKer1}) and Eq.~(\ref{IntKer2}) gives us:%
\begin{align}
K_{R}(\mathbf{x}_{1},t_{1};\mathbf{x}_{2},t_{2})  &  =K_{L}^{\ast}%
(\mathbf{x}_{1},t_{1};\mathbf{x}_{2},t_{2}),\nonumber\\
K_{L}(\mathbf{x}_{1},t_{1};\mathbf{x}_{2},t_{2})  &  =K_{R}^{\ast}%
(\mathbf{x}_{1},t_{1};\mathbf{x}_{2},t_{2}). \label{ZusProSch}%
\end{align}

The propagators must satisfy the principle of causality to describe the time
evolution of the Schr\"{o}\-dinger wave functions correctly. The wave function
at time $t$ cannot depend on the wave function at later times $t^{\hspace
{0.01in}\prime}>t$. The \textbf{retarded propagators} satisfy this
requirement:%
\begin{align}
(K_{R})^{+}(\mathbf{x}^{\hspace{0.01in}\prime}\hspace{-0.01in},t^{\hspace
{0.01in}\prime}\hspace{-0.01in};\mathbf{x},t)  &  =\theta(t^{\hspace
{0.01in}\prime}\hspace{-0.02in}-t)\,K_{R}(\mathbf{x}^{\hspace{0.01in}\prime
}\hspace{-0.01in},t^{\hspace{0.01in}\prime}\hspace{-0.01in};\mathbf{x}%
,t),\nonumber\\
(K_{L})^{+}(\mathbf{x},t;\mathbf{x}^{\hspace{0.01in}\prime}\hspace
{-0.01in},t^{\hspace{0.01in}\prime})  &  =\theta(t^{\hspace{0.01in}\prime
}\hspace{-0.02in}-t)\,K_{L}(\mathbf{x},t;\mathbf{x}^{\hspace{0.01in}\prime
}\hspace{-0.01in},t^{\hspace{0.01in}\prime}),\label{DefRetPro1}\\[0.1in]
(K_{R}^{\ast})^{+}(\mathbf{x}^{\hspace{0.01in}\prime}\hspace{-0.01in}%
,t^{\hspace{0.01in}\prime}\hspace{-0.01in};\mathbf{x},t)  &  =\theta
(t^{\hspace{0.01in}\prime}\hspace{-0.02in}-t)\,K_{R}^{\ast}(\mathbf{x}%
^{\hspace{0.01in}\prime}\hspace{-0.01in},t^{\hspace{0.01in}\prime}%
\hspace{-0.01in};\mathbf{x},t),\nonumber\\
(K_{L}^{\ast})^{+}(\mathbf{x},t;\mathbf{x}^{\hspace{0.01in}\prime}%
\hspace{-0.01in},t^{\hspace{0.01in}\prime})  &  =\theta(t^{\hspace
{0.01in}\prime}\hspace{-0.02in}-t)\,K_{L}^{\ast}(\mathbf{x},t;\mathbf{x}%
^{\hspace{0.01in}\prime}\hspace{-0.01in},t^{\hspace{0.01in}\prime}).
\label{DefRetPro2}%
\end{align}
Note that $\theta(t)$ stands for the Heaviside function:%
\begin{equation}
\theta(t)=%
\begin{cases}
1 & \text{if }t\geq0,\\
0 & \text{otherwise}.
\end{cases}
\end{equation}
The \textbf{advanced propagators}, on the other hand, describe the propagation
of a wave function backward in time:%
\begin{align}
(K_{R})^{-}(\mathbf{x}^{\hspace{0.01in}\prime},t^{\hspace{0.01in}\prime
}\hspace{-0.01in};\mathbf{x},t)  &  =\theta(t-t^{\hspace{0.01in}\prime
})\,K_{R}(\mathbf{x}^{\hspace{0.01in}\prime},t^{\hspace{0.01in}\prime}%
\hspace{-0.01in};\mathbf{x},t),\nonumber\\
(K_{L})^{-}(\mathbf{x},t;\mathbf{x}^{\hspace{0.01in}\prime},t^{\hspace
{0.01in}\prime})  &  =\theta(t-t^{\hspace{0.01in}\prime})\,K_{L}%
(\mathbf{x},t;\mathbf{x}^{\hspace{0.01in}\prime},t^{\hspace{0.01in}\prime
}),\label{DefAdvPro1}\\[0.1in]
(K_{R}^{\ast})^{-}(\mathbf{x}^{\hspace{0.01in}\prime},t^{\hspace{0.01in}%
\prime}\hspace{-0.01in};\mathbf{x},t)  &  =\theta(t-t^{\hspace{0.01in}\prime
})\,K_{R}^{\ast}(\mathbf{x}^{\hspace{0.01in}\prime},t^{\hspace{0.01in}\prime
}\hspace{-0.01in};\mathbf{x},t),\nonumber\\
(K_{L}^{\ast})^{-}(\mathbf{x},t;\mathbf{x}^{\hspace{0.01in}\prime}%
\hspace{-0.01in},t^{\hspace{0.01in}\prime})  &  =\theta(t-t^{\hspace
{0.01in}\prime})\,K_{L}^{\ast}(\mathbf{x},t;\mathbf{x}^{\hspace{0.01in}\prime
}\hspace{-0.01in},t^{\hspace{0.01in}\prime}). \label{DefAdvPro2}%
\end{align}
From Eq.~(\ref{ZusProSch}) follows:%
\begin{align}
(K_{R})^{\pm}(\mathbf{x}_{1},t_{1};\mathbf{x}_{2},t_{2})  &  =(K_{L}^{\ast
})^{\mp}(\mathbf{x}_{1},t_{1};\mathbf{x}_{2},t_{2}),\nonumber\\
(K_{L})^{\pm}(\mathbf{x}_{1},t_{1};\mathbf{x}_{2},t_{2})  &  =(K_{R}^{\ast
})^{\mp}(\mathbf{x}_{1},t_{1};\mathbf{x}_{2},t_{2}). \label{ZusGreAvaRetN}%
\end{align}

The propagators in Eqs.~(\ref{DefRetPro1}) and (\ref{DefRetPro2}) are
solutions to inhomogeneous wave equations. We show this for the propagator
$(K_{R})^{+}(\mathbf{x}^{\hspace{0.01in}\prime},t^{\hspace{0.01in}\prime
};\mathbf{x},t)$:%
\begin{align}
&  \text{i}\partial_{t^{\prime}}\triangleright(K_{R})^{+}(\mathbf{x}%
^{\hspace{0.01in}\prime}\hspace{-0.01in},t^{\hspace{0.01in}\prime}%
\hspace{-0.01in};\mathbf{x},t)=\nonumber\\
&  \qquad=\Big (\text{i}\frac{\partial}{\partial\hspace{0.01in}t^{\hspace
{0.01in}\prime}}\theta(t^{\hspace{0.01in}\prime}\hspace{-0.02in}%
-t)\Big )\,K_{R}(\mathbf{x}^{\hspace{0.01in}\prime}\hspace{-0.01in}%
,t^{\hspace{0.01in}\prime}\hspace{-0.01in};\mathbf{x},t)+\theta(t^{\hspace
{0.01in}\prime}\hspace{-0.02in}-t)\,\text{i}\frac{\partial}{\partial
\hspace{0.01in}t^{\hspace{0.01in}\prime}}K_{R}(\mathbf{x}^{\hspace
{0.01in}\prime}\hspace{-0.01in},t^{\hspace{0.01in}\prime}\hspace
{-0.01in};\mathbf{x},t)\nonumber\\
&  \qquad=\,\text{i\hspace{0.01in}}\delta(t^{\hspace{0.01in}\prime}%
\hspace{-0.02in}-t)\,K_{R}(\mathbf{x}^{\hspace{0.01in}\prime}\hspace
{-0.01in},t^{\hspace{0.01in}\prime}\hspace{-0.01in};\mathbf{x},t)+\theta
(t^{\hspace{0.01in}\prime}\hspace{-0.02in}-t)\,H_{0}^{\prime}\triangleright
K_{R}(\mathbf{x}^{\hspace{0.01in}\prime},t^{\hspace{0.01in}\prime}%
;\mathbf{x},t)\nonumber\\[0.03in]
&  \qquad=\,\text{i}\operatorname*{vol}\nolimits^{-1}\hspace{-0.01in}%
\delta(t^{\hspace{0.01in}\prime}\hspace{-0.02in}-t)\,\delta_{q}^{\hspace{0.01in}3}%
(\mathbf{x}^{\hspace{0.01in}\prime}\oplus(\ominus\hspace{0.01in}\kappa
^{-1}\mathbf{x}))+H_{0}^{\prime}\triangleright(K_{R})^{+}(\mathbf{x}%
^{\hspace{0.01in}\prime},t^{\hspace{0.01in}\prime};\mathbf{x},t).
\label{SchrGlGreHab}%
\end{align}
In the second step of the calculation above, we have taken into account that
applying the time derivative to the Heaviside function gives the classical
delta function. Moreover, we have used the result of the following calculation
[see Eq.~(\ref{IntKer1}) as well as Eq.~(\ref{FreSchGlImp0}) of
Chap.~\ref{LoeSchGleKap}]%
\begin{align}
\text{i}\frac{\partial}{\partial\hspace{0.01in}t^{\hspace{0.01in}\prime}}%
K_{R}(\mathbf{x}^{\hspace{0.01in}\prime}\hspace{-0.01in},t^{\hspace
{0.01in}\prime}\hspace{-0.01in};\mathbf{x},t)  &  =\int\text{d}_{q}%
^{3}p\,\frac{\partial}{\partial\hspace{0.01in}t^{\hspace{0.01in}\prime}%
}u_{\hspace{0.01in}\mathbf{p}}(\mathbf{x}^{\hspace{0.01in}\prime}%
,t^{\hspace{0.01in}\prime})\circledast(u^{\ast})_{\mathbf{p}}(\mathbf{x}%
,t)\nonumber\\
&  =\int\text{d}_{q}^{3}p\,H_{0}\overset{x^{\prime}}{\triangleright}%
u_{\hspace{0.01in}\mathbf{p}}(\mathbf{x}^{\hspace{0.01in}\prime}%
,t^{\hspace{0.01in}\prime})\circledast(u^{\ast})_{\mathbf{p}}(\mathbf{x}%
,t)\nonumber\\
&  =H_{0}\overset{x^{\prime}}{\triangleright}K_{R}(\mathbf{x}^{\hspace
{0.01in}\prime}\hspace{-0.01in},t^{\hspace{0.01in}\prime}\hspace
{-0.01in};\mathbf{x},t).
\end{align}
Note that the last step in Eq.~(\ref{SchrGlGreHab}) follows from the following
identities [cf. Eq.~(\ref{VolRelZeiWelDreDim1}) of the previous chapter]:%
\begin{align}
\lim_{t^{\prime}\rightarrow\hspace{0.01in}t}\,K_{R}(\mathbf{x}^{\hspace
{0.01in}\prime}\hspace{-0.01in},t^{\hspace{0.01in}\prime}\hspace
{-0.01in};\mathbf{x},t)  &  =\int\text{d}_{q}^{3}p\,u_{\hspace{0.01in}%
\mathbf{p}}(\mathbf{x}^{\hspace{0.01in}\prime})\circledast(u^{\ast
})_{\mathbf{p}}(\mathbf{x})\nonumber\\
&  =\operatorname*{vol}\nolimits^{-1}\hspace{-0.01in}\delta_{q}^{\hspace{0.01in}3}%
(\mathbf{x}^{\hspace{0.01in}\prime}\oplus(\ominus\hspace{0.01in}\kappa
^{-1}\mathbf{x})). \label{RanBedKLSch}%
\end{align}
The same arguments yield the following result for the advanced propagator:%
\begin{equation}
(\text{i}\partial_{t^{\prime}}-H_{0}^{\prime})\triangleright(K_{R}%
)^{-}(\mathbf{x}^{\hspace{0.01in}\prime},t^{\hspace{0.01in}\prime}%
;\mathbf{x},t)=-\,\text{i}\operatorname*{vol}\nolimits^{-1}\hspace
{-0.01in}\delta(t^{\hspace{0.01in}\prime}\hspace{-0.02in}-t)\,\delta_{q}%
^{\hspace{0.01in}3}(\mathbf{x}^{\hspace{0.01in}\prime}\oplus(\ominus\hspace{0.01in}%
\kappa^{-1}\mathbf{x})). \label{SchrGlGreHab2}%
\end{equation}
The other $q$-de\-formed versions of the Schr\"{o}\-dinger propagator satisfy similar wave
equations, i.~e.%
\begin{equation}
(K_{L})^{\pm}(\mathbf{x},t;\mathbf{x}^{\hspace{0.01in}\prime}\hspace
{-0.01in},t^{\hspace{0.01in}\prime})\,\bar{\triangleleft}\,(\partial
_{t^{\prime}}\text{i}-H_{0}^{\prime})=\mp\,\text{i}\operatorname*{vol}%
\nolimits^{-1}\hspace{-0.01in}\delta(t-t^{\hspace{0.01in}\prime})\,\delta
_{q}^{\hspace{0.01in}3}((\ominus\hspace{0.01in}\kappa^{-1}\mathbf{x})\oplus\mathbf{x}%
^{\hspace{0.01in}\prime})
\end{equation}
or%
\begin{align}
(\text{i}\partial_{t^{\prime}}-H_{0}^{\prime})\,\bar{\triangleright}%
\,(K_{R}^{\ast})^{\pm}(\mathbf{x}^{\hspace{0.01in}\prime}\hspace
{-0.01in},t^{\hspace{0.01in}\prime}\hspace{-0.01in};\mathbf{x},t)  &
=\pm\,\text{i}\operatorname*{vol}\nolimits^{-1}\hspace{-0.01in}\delta
(t^{\hspace{0.01in}\prime}\hspace{-0.02in}-t)\,\delta_{q}^{\hspace{0.01in}3}((\ominus
\hspace{0.01in}\kappa^{-1}\mathbf{x}^{\hspace{0.01in}\prime})\oplus
\mathbf{x}),\nonumber\\
(K_{L}^{\ast})^{\pm}(\mathbf{x},t;\mathbf{x}^{\hspace{0.01in}\prime}%
\hspace{-0.01in},t^{\hspace{0.01in}\prime})\triangleleft(\partial_{t^{\prime}%
}\text{i}-H_{0}^{\prime})  &  =\mp\,\text{i}\operatorname*{vol}\nolimits^{-1}%
\hspace{-0.01in}\delta(t-t^{\hspace{0.01in}\prime})\,\delta_{q}^{\hspace{0.01in}3}%
(\mathbf{x}\oplus(\ominus\hspace{0.01in}\kappa^{-1}\mathbf{x}^{\hspace
{0.01in}\prime})). \label{SchrGlGreHabEnd}%
\end{align}

We can use our $q$-de\-formed Schr\"{o}\-dinger propagators to get solutions to the following
inhomogeneous wave equations:%
\begin{align}
(\text{i}\partial_{t}-H_{0})\triangleright(\psi_{R})^{\pm}(\mathbf{x},t)  &
=\varrho(\mathbf{x},t),\nonumber\\
(\psi_{L})^{\pm}(\mathbf{x},t)\,\bar{\triangleleft}\,(\partial_{t}%
\text{i\hspace{0.01in}}-H_{0})  &  =\varrho(\mathbf{x}%
,t),\label{InhSchrGleAll}\\[0.1in]
(\text{i}\partial_{t}-H_{0})\,\bar{\triangleright}\,(\psi_{R}^{\ast})^{\pm
}(\mathbf{x},t)  &  =\varrho(\mathbf{x},t),\nonumber\\
(\psi_{L}^{\ast})^{\pm}(\mathbf{x},t)\triangleleft(\partial_{t}\text{i\hspace
{0.01in}}-H_{0})  &  =\varrho(\mathbf{x},t).
\end{align}
Due Eq.~(\ref{SchrGlGreHab}) and Eqs.~(\ref{SchrGlGreHab2}%
)-(\ref{SchrGlGreHabEnd}), these solutions are%
\begin{align}
(\psi_{R})^{\pm}(\mathbf{x},t)  &  =\mp\,\text{i}\int\text{d}t^{\hspace
{0.01in}\prime}\hspace{-0.01in}\int\text{d}_{q}^{3}x^{\prime}\,(K_{R})^{\pm
}(\mathbf{x},t;\mathbf{x}^{\hspace{0.01in}\prime}\hspace{-0.01in}%
,t^{\hspace{0.01in}\prime})\circledast\varrho(\mathbf{x}^{\hspace
{0.01in}\prime}\hspace{-0.01in},t^{\hspace{0.01in}\prime}),\nonumber\\
(\psi_{L})^{\pm}(\mathbf{x},t)  &  =\pm\,\text{i}\int\text{d}t^{\hspace
{0.01in}\prime}\hspace{-0.01in}\int\text{d}_{q}^{3}x^{\prime}\,\varrho
(\mathbf{x}^{\hspace{0.01in}\prime},t^{\hspace{0.01in}\prime})\circledast
(K_{L})^{\pm}(\mathbf{x}^{\hspace{0.01in}\prime}\hspace{-0.01in}%
,t^{\hspace{0.01in}\prime}\hspace{-0.01in};\mathbf{x}%
,t),\label{SolInhSchr1Hab}%
\end{align}
and
\begin{align}
(\psi_{R}^{\ast})^{\pm}(\mathbf{x},t)  &  =\mp\,\text{i}\int\text{d}%
t^{\hspace{0.01in}\prime}\hspace{-0.01in}\int\text{d}_{q}^{3}x^{\prime
}\,(K_{R}^{\ast})^{\pm}(\mathbf{x},t;\mathbf{x}^{\hspace{0.01in}\prime}%
\hspace{-0.01in},t^{\hspace{0.01in}\prime})\circledast\varrho(\mathbf{x}%
^{\hspace{0.01in}\prime}\hspace{-0.01in},t^{\hspace{0.01in}\prime
}),\nonumber\\
(\psi_{L}^{\ast})^{\pm}(\mathbf{x},t)  &  =\pm\,\text{i}\int\text{d}%
t^{\hspace{0.01in}\prime}\hspace{-0.01in}\int\text{d}_{q}^{3}x^{\prime
}\,\varrho(\mathbf{x}^{\hspace{0.01in}\prime}\hspace{-0.01in},t^{\hspace
{0.01in}\prime})\circledast(K_{L}^{\ast})^{\pm}(\mathbf{x}^{\hspace
{0.01in}\prime}\hspace{-0.01in},t^{\hspace{0.01in}\prime}\hspace
{-0.01in};\mathbf{x},t). \label{SolInhSchr2Hab}%
\end{align}
By way of example, we check that the expression for $(\psi_{R})^{\pm
}(\mathbf{x},t)$ satisfies the first identity in\ Eq.~(\ref{InhSchrGleAll}):%
\begin{align}
&  (\text{i}\partial_{t}-H_{0})\triangleright(\psi_{R})^{\pm}(\mathbf{x}%
,t)=\nonumber\\
&  \qquad=\mp\,\text{i}\int\text{d}t^{\hspace{0.01in}\prime}\int\text{d}%
_{q}^{3}x^{\prime}\,(\text{i}\partial_{t}-H_{0})\triangleright(K_{R})^{\pm
}(\mathbf{x},t;\mathbf{x}^{\hspace{0.01in}\prime}\hspace{-0.01in}%
,t^{\hspace{0.01in}\prime})\circledast\varrho(\mathbf{x}^{\hspace
{0.01in}\prime}\hspace{-0.01in},t^{\hspace{0.01in}\prime})\nonumber\\
&  \qquad=\operatorname*{vol}\nolimits^{-1}\hspace{-0.02in}\int\text{d}%
t^{\hspace{0.01in}\prime}\,\delta(t-t^{\hspace{0.01in}\prime})\int\text{d}%
_{q}^{3}x^{\prime}\,\delta_{q}^{\hspace{0.01in}3}(\mathbf{x}\oplus(\ominus\hspace
{0.01in}\kappa^{-1}\mathbf{x}^{\hspace{0.01in}\prime}))\circledast
\varrho(\mathbf{x}^{\hspace{0.01in}\prime}\hspace{-0.01in},t^{\hspace
{0.01in}\prime})\nonumber\\
&  \qquad=\operatorname*{vol}\nolimits^{-1}\hspace{-0.02in}\int\text{d}%
_{q}^{3}x^{\prime}\,\delta_{q}^{\hspace{0.01in}3}(\mathbf{x}\oplus(\ominus\hspace
{0.01in}\kappa^{-1}\mathbf{x}^{\hspace{0.01in}\prime}))\circledast
\varrho(\mathbf{x}^{\hspace{0.01in}\prime}\hspace{-0.01in},t)=\varrho
(\mathbf{x},t).
\end{align}
Note that the second step of the above calculation results from
Eq.~(\ref{SchrGlGreHab}). In the last step, we made use of the identities
given in Eq.~(\ref{AlgChaIdeqDelFkt}) of Chap.~\ref{KapOrtVolEBeWel}.

Next, we derive some useful identities for the $q$-de\-formed Schr\"{o}\-dinger propagators. We
multiply both sides of the integral equations given in
Eqs.~(\ref{DefProDreDim1}) and (\ref{DefProDreDim2}) by the Heaviside function
and take into account Eqs.~(\ref{DefRetPro1}) and (\ref{DefRetPro2}). Thus, we
get%
\begin{align}
\theta(t^{\hspace{0.01in}\prime}\hspace{-0.02in}-t)\,\phi_{R}(\mathbf{x}%
^{\hspace{0.01in}\prime}\hspace{-0.01in},t^{\hspace{0.01in}\prime})  &
=\int\text{d}_{q}^{3}x\,(K_{R})^{+}(\mathbf{x}^{\hspace{0.01in}\prime}%
\hspace{-0.01in},t^{\hspace{0.01in}\prime}\hspace{-0.01in};\mathbf{x}%
,t)\circledast\phi_{R}(\mathbf{x},t),\nonumber\\
\theta(t^{\hspace{0.01in}\prime}\hspace{-0.02in}-t)\,\phi_{L}(\mathbf{x}%
^{\hspace{0.01in}\prime}\hspace{-0.01in},t^{\hspace{0.01in}\prime})  &
=\int\text{d}_{q}^{3}x\,\phi_{L}(\mathbf{x},t)\circledast(K_{L})^{+}%
(\mathbf{x},t;\mathbf{x}^{\hspace{0.01in}\prime}\hspace{-0.01in}%
,t^{\hspace{0.01in}\prime}), \label{ChaGleSchPro1}%
\end{align}
and%
\begin{align}
\theta(t^{\hspace{0.01in}\prime}\hspace{-0.02in}-t)\,\phi_{R}^{\ast
}(\mathbf{x}^{\hspace{0.01in}\prime}\hspace{-0.01in},t^{\hspace{0.01in}\prime
})  &  =\int\text{d}_{q}^{3}x\,(K_{R}^{\ast})^{+}(\mathbf{x}^{\hspace
{0.01in}\prime}\hspace{-0.01in},t^{\hspace{0.01in}\prime}\hspace
{-0.01in};\mathbf{x},t)\circledast\phi_{R}^{\ast}(\mathbf{x},t),\nonumber\\
\theta(t^{\hspace{0.01in}\prime}\hspace{-0.02in}-t)\,\phi_{L}^{\ast
}(\mathbf{x}^{\hspace{0.01in}\prime}\hspace{-0.01in},t^{\hspace{0.01in}\prime
})  &  =\int\text{d}_{q}^{3}x\,\phi_{L}^{\ast}(\mathbf{x},t)\circledast
(K_{L}^{\ast})^{+}(\mathbf{x},t;\mathbf{x}^{\hspace{0.01in}\prime}%
\hspace{-0.01in},t^{\hspace{0.01in}\prime}). \label{ChaGleSchPro2}%
\end{align}
Similar relations hold for the advanced propagators. Applying the first
identity of Eq.~(\ref{ChaGleSchPro1}) twice, we obtain:%
\begin{align}
\phi_{R}(\mathbf{x}^{\hspace{0.01in}\prime}\hspace{-0.01in},t^{\hspace
{0.01in}\prime})=  &  \int\text{d}_{q}^{3}x\,(K_{R})^{+}(\mathbf{x}%
^{\hspace{0.01in}\prime}\hspace{-0.01in},t^{\hspace{0.01in}\prime}%
\hspace{-0.01in};\mathbf{x},t)\circledast\phi_{R}(\mathbf{x},t)\nonumber\\
=  &  \int\text{d}_{q}^{3}x\int\text{d}_{q}^{3}x^{\hspace{0.01in}\prime\prime
}(K_{R})^{+}(\mathbf{x}^{\hspace{0.01in}\prime}\hspace{-0.01in},t^{\hspace
{0.01in}\prime}\hspace{-0.01in};\mathbf{x}^{\hspace{0.01in}\prime\prime
}\hspace{-0.01in},t^{\hspace{0.01in}\prime\prime})\nonumber\\
&  \circledast(K_{R})^{+}(\mathbf{x}^{\hspace{0.01in}\prime\prime}%
\hspace{-0.01in},t^{\hspace{0.01in}\prime\prime}\hspace{-0.01in}%
;\mathbf{x},t)\circledast\phi_{R}(\mathbf{x},t). \label{HerZusPro}%
\end{align}
Since we have assumed $t\leq t^{\hspace{0.01in}\prime\prime}\leq
t^{\hspace{0.01in}\prime}$ for the retarded propagators, we could omit the Heaviside functions in the
expressions above. By comparing the second expression in Eq.~(\ref{HerZusPro})
to the last one, we can see that the following identity holds:\footnote{For the advanced propagators, we have  $t^{\hspace{0.01in}\prime}\leq t^{\hspace{0.01in}\prime\prime}\leq
t$. }%
\begin{equation}
(K_{R})^{\pm}(\mathbf{x}^{\hspace{0.01in}\prime}\hspace{-0.01in}%
,t^{\hspace{0.01in}\prime}\hspace{-0.01in};\mathbf{x},t)=\int\text{d}_{q}%
^{3}x^{\hspace{0.01in}\prime\prime}(K_{R})^{\pm}(\mathbf{x}^{\hspace
{0.01in}\prime}\hspace{-0.01in},t^{\hspace{0.01in}\prime}\hspace
{-0.01in};\mathbf{x}^{\hspace{0.01in}\prime\prime}\hspace{-0.01in}%
,t^{\hspace{0.01in}\prime\prime})\circledast(K_{R})^{\pm}(\mathbf{x}%
^{\hspace{0.01in}\prime\prime}\hspace{-0.01in},t^{\hspace{0.01in}\prime\prime
}\hspace{-0.01in};\mathbf{x},t). \label{ComFree1Hab}%
\end{equation}
In the same manner, we get:%
\begin{equation}
(K_{L})^{\pm}(\mathbf{x},t;\mathbf{x}^{\hspace{0.01in}\prime}\hspace
{-0.01in},t^{\hspace{0.01in}\prime})=\int\text{d}_{q}^{3}x^{\hspace
{0.01in}\prime\prime}(K_{L})^{\pm}(\mathbf{x},t;\mathbf{x}^{\hspace
{0.01in}\prime\prime}\hspace{-0.01in},t^{\hspace{0.01in}\prime\prime
})\circledast(K_{L})^{\pm}(\mathbf{x}^{\hspace{0.01in}\prime\prime}%
\hspace{-0.01in},t^{\hspace{0.01in}\prime\prime}\hspace{-0.01in}%
;\mathbf{x}^{\hspace{0.01in}\prime}\hspace{-0.01in},t^{\hspace{0.01in}\prime
}).
\end{equation}

We can also show how the $q$-de\-formed Schr\"{o}\-dinger propagators behave under conjugation.
From the conjugation properties of the $q$-de\-formed plane waves [cf.
Eq.~(\ref{KonEbeWel}) in Chap.~\ref{LoeSchGleKap}] together with the formulas
given in Eqs.~(\ref{IntKer1}) and (\ref{IntKer2}) follows:%
\begin{align}
\overline{(K_{R})^{\pm}(\mathbf{x}^{\hspace{0.01in}\prime}\hspace
{-0.01in},t^{\hspace{0.01in}\prime}\hspace{-0.01in};\mathbf{x},t)}  &
=(K_{L})^{\pm}(\mathbf{x},t;\mathbf{x}^{\hspace{0.01in}\prime}\hspace
{-0.01in},t^{\hspace{0.01in}\prime}),\nonumber\\
(\overline{K_{L}^{\ast})^{\pm}(\mathbf{x},t;\mathbf{x}^{\hspace{0.01in}\prime
}\hspace{-0.01in},t^{\hspace{0.01in}\prime})}  &  =(K_{R}^{\ast})^{\pm
}(\mathbf{x}^{\hspace{0.01in}\prime}\hspace{-0.01in},t^{\hspace{0.01in}\prime
}\hspace{-0.01in};\mathbf{x},t). \label{KonFreProNR}%
\end{align}

Finally, we show how to derive the momentum space form of the $q$-de\-formed Schr\"{o}\-dinger
propagators. We demonstrate our considerations using the example of the
propagator $(K_{R})^{\pm}(\mathbf{x}^{\hspace{0.01in}\prime}\hspace
{-0.01in},t^{\hspace{0.01in}\prime}\hspace{-0.01in};\mathbf{x},t)$:%
\begin{align}
&  (K_{R})^{\pm}(\mathbf{x}^{\hspace{0.01in}\prime}\hspace{-0.01in}%
,t^{\hspace{0.01in}\prime}\hspace{-0.01in};\mathbf{x},t)=\theta(\pm
(t^{\hspace{0.01in}\prime}\hspace{-0.02in}-t))\int\text{d}_{q}^{3}%
p\,u_{\hspace{0.01in}\mathbf{p}}(\mathbf{x}^{\hspace{0.01in}\prime}%
,t^{\hspace{0.01in}\prime})\circledast(u^{\ast})_{\mathbf{p}}(\mathbf{x}%
,t)\nonumber\\
&  =\frac{\pm\hspace{0.01in}\text{i}}{2\pi}\lim_{\varepsilon\hspace
{0.01in}\rightarrow\hspace{0.01in}0^{+}}\int\nolimits_{-\infty}^{+\infty
}\text{d}E\,\frac{\operatorname{e}^{\text{i}E(t-t^{\prime})}}{E\pm
\text{i}\varepsilon}\int\text{d}_{q}^{3}p\,u_{\hspace{0.01in}\mathbf{p}%
}(\mathbf{x}^{\hspace{0.01in}\prime})\circledast\exp\left(  \frac
{\text{i}\mathbf{p}^{2}(t-t^{\hspace{0.01in}\prime})}{2\hspace{0.01in}%
m}\right)  \circledast(u^{\ast})_{\mathbf{p}}(\mathbf{x})\nonumber\\
&  =\frac{1}{2\pi}\int\text{d}_{q}^{3}p\,u_{\hspace{0.01in}\mathbf{p}%
}(\mathbf{x}^{\hspace{0.01in}\prime})\circledast\left(  \int\text{d}%
E\operatorname{e}^{\text{i}E(t-t^{\prime})}(K_{R})^{\pm}(\mathbf{p},E)\right)
\circledast(u^{\ast})_{\mathbf{p}}(\mathbf{x}). \label{GreFktImpRau}%
\end{align}
In the first step of the above calculation, we wrote the Heaviside function as an integral. Next, we replaced the energy variable $E$ by
$E-\mathbf{p}^{2}/(2\hspace{0.01in}m)$. The later is possible because
$\mathbf{p}^{2}$ is a central element of the momentum algebra. This way, we can read off the
Schr\"{o}\-din\-ger propagator in momentum space:%
\begin{equation}
(K_{R})^{\pm}(\mathbf{p},E)=\pm\hspace{0.01in}\text{i}(E-\mathbf{p}%
^{2}/(2\hspace{0.01in}m)\pm\,\text{i}\varepsilon)^{-1}. \label{SchProImpL}%
\end{equation}
Note that we must write the Schr\"{o}\-dinger propagator in momentum space as a
series in nor\-mal-or\-dered monomials of momentum coordinates. To get this
series, we first write the right-hand side of Eq.~(\ref{SchProImpL}) as a
power series of $\mathbf{p}^{2}$. Then we apply the formula in
Eq.~(\ref{EntPotP}) of Chap.~\ref{LoeSchGleKap}.

Similar reasonings hold for the other $q$-de\-formed versions of the Schr\"{o}\-dinger
propagator. Thus we also have%
\begin{gather}
(K_{L})^{\pm}(\mathbf{x},t;\mathbf{x}^{\hspace{0.01in}\prime}\hspace
{-0.01in},t^{\hspace{0.01in}\prime})=\theta(\pm(t^{\hspace{0.01in}\prime
}\hspace{-0.02in}-t))\int\text{d}_{q}^{3}p\,(u^{\ast})^{\mathbf{p}}%
(\mathbf{x},t)\circledast u^{\mathbf{p}}(\mathbf{x}^{\hspace{0.01in}\prime
},t^{\hspace{0.01in}\prime})\nonumber\\
=\frac{1}{2\pi}\int\text{d}_{q}^{3}p\,(u^{\ast})^{\mathbf{p}}(\mathbf{x}%
)\circledast\left(  \int\text{d}E\operatorname{e}^{\text{i}E(t-t^{\prime}%
)}(K_{L})^{\pm}(\mathbf{p},E)\right)  \circledast u^{\mathbf{p}}%
(\mathbf{x}^{\hspace{0.01in}\prime}), \label{GreFktImpRau2}%
\end{gather}
where%
\begin{equation}
(K_{L})^{\pm}(\mathbf{p},E)=\pm\hspace{0.01in}\text{i}(E+\mathbf{p}%
^{2}/(2\hspace{0.01in}m)\pm\text{i}\varepsilon)^{-1}. \label{SchProImpR}%
\end{equation}
Taking into account Eq.~(\ref{ZusGreAvaRetN}), we also have%
\begin{align}
&  (K_{R}^{\ast})^{\pm}(\mathbf{x}^{\hspace{0.01in}\prime}\hspace
{-0.01in},t^{\hspace{0.01in}\prime}\hspace{-0.01in};\mathbf{x},t)=\nonumber\\
&  \qquad=\frac{1}{2\pi}\int\text{d}_{q}^{3}p\,(u^{\ast})^{\mathbf{p}%
}(\mathbf{x}^{\hspace{0.01in}\prime})\circledast\left(  \int\text{d}%
E\operatorname{e}^{\text{i}E(t-t^{\prime})}(K_{R}^{\ast})^{\pm}(\mathbf{p}%
,E)\right)  \circledast u^{\mathbf{p}}(\mathbf{x}), \label{GreFktImpRau3}%
\\[0.08in]
&  (K_{L}^{\ast})^{\pm}(\mathbf{x},t;\mathbf{x}^{\hspace{0.01in}\prime}%
\hspace{-0.01in},t^{\hspace{0.01in}\prime})=\nonumber\\
&  \qquad=\frac{1}{2\pi}\int\text{d}_{q}^{3}p\,u_{\hspace{0.01in}\mathbf{p}%
}(\mathbf{x})\circledast\left(  \int\text{d}E\operatorname{e}^{\text{i}%
E(t-t^{\prime})}(K_{L}^{\ast})^{\pm}(\mathbf{p},E)\right)  \circledast
(u^{\ast})_{\mathbf{p}}(\mathbf{x}^{\hspace{0.01in}\prime})
\label{GreFktImpRau4}%
\end{align}
with%
\begin{align}
(K_{R}^{\ast})^{\pm}(\mathbf{p},E)  &  =\pm\hspace{0.01in}\text{i}%
(E-\mathbf{p}^{2}/(2\hspace{0.01in}m)\pm\text{i}\varepsilon)^{-1},\nonumber\\
(K_{L}^{\ast})^{\pm}(\mathbf{p},E)  &  =\pm\hspace{0.01in}\text{i}%
(E+\mathbf{p}^{2}/(2\hspace{0.01in}m)\pm\text{i}\varepsilon)^{-1}.
\end{align}

\begin{figure}
[ptb]
\centerline{\psfig{figure=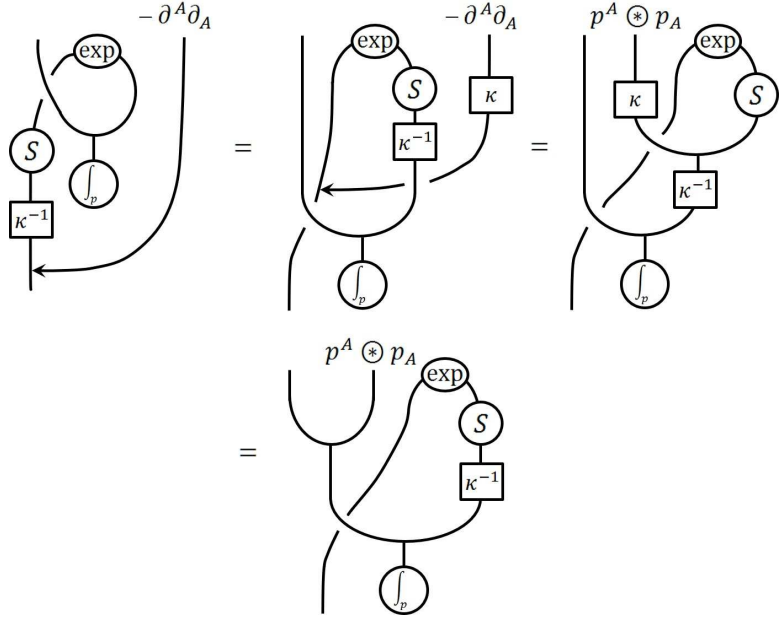,width=4.0257in}}%
\caption{Proof for the first step in Eq.~(\ref{DifGleProSchrGle}).}%
\label{Bild82}%
\end{figure}

Immediately, we can verify that the expressions in Eqs.~(\ref{GreFktImpRau}),
(\ref{GreFktImpRau2}), (\ref{GreFktImpRau3}), and (\ref{GreFktImpRau4})
satisfy the wave equations given in Eqs.~(\ref{SchrGlGreHab}) and
(\ref{SchrGlGreHab2})-(\ref{SchrGlGreHabEnd}). The following applies, for
example:%
\begin{align}
&  (K_{L}^{\ast})^{\pm}(\mathbf{x},t;\mathbf{x}^{\hspace{0.01in}\prime}%
\hspace{-0.01in},t^{\hspace{0.01in}\prime})\triangleleft(\partial_{t^{\prime}%
}\text{i}-H_{0}^{\prime})=\nonumber\\
&  \qquad=\frac{\pm\hspace{0.01in}\text{i}}{2\pi}\int\text{d}_{q}%
^{3}p\,u_{\hspace{0.01in}\mathbf{p}}(\mathbf{x})\circledast\lim_{\varepsilon
\hspace{0.01in}\rightarrow\hspace{0.01in}0^{+}}\int\text{d}E\operatorname{e}%
^{\text{i}E(t-t^{\prime})}(E+\mathbf{p}^{2}/(2\hspace{0.01in}m)\pm
\text{i}\varepsilon)^{-1}\nonumber\\
&  \qquad\qquad\qquad\qquad\qquad\circledast(-E-\mathbf{p}^{2}/(2\hspace
{0.01in}m))\circledast(u^{\ast})_{\mathbf{p}}(\mathbf{x}^{\hspace
{0.01in}\prime})\nonumber\\
&  \qquad=\frac{\mp\hspace{0.01in}\text{i}}{2\pi}\int\text{d}%
E\,\operatorname{e}^{\text{i}E(t-t^{\prime})}\hspace{-0.01in}\int\text{d}%
_{q}^{3}p\,u_{\hspace{0.01in}\mathbf{p}}(\mathbf{x})\circledast(u^{\ast
})_{\mathbf{p}}(\mathbf{x}^{\hspace{0.01in}\prime})\nonumber\\
&  \qquad=\mp\,\text{i}\operatorname*{vol}\nolimits^{-1}\hspace{-0.01in}%
\delta(t-t^{\hspace{0.01in}\prime})\,\delta_{q}^{\hspace{0.01in}3}(\mathbf{x}\oplus
(\ominus\hspace{0.01in}\kappa^{-1}\mathbf{x}^{\hspace{0.01in}\prime
})).\label{DifGleProSchrGle}%
\end{align}
In the first step of the calculation in Eq.~(\ref{DifGleProSchrGle}), we made
use of an identity that we have proven in Fig.~\ref{Bild82} by graphical
methods.\footnote{How to apply these graphical methods see
Ref.~\cite{Majid:2002kd} and the appendix of  Ref.~\cite{Wachter:2019A}.} The
last step in Eq.~(\ref{DifGleProSchrGle}) follows from the completeness
relations in Eq.~(\ref{VolRelZeiWelDreDim1}) of
Chap.~\ref{KapOrtVolEBeWel} by setting $t=0$.%

\section{Expectation values of position or momentum\label{ErwOrtImpKapN}}

In this chapter, we consider the expectation values of the operators for
momentum and position. We calculate these expectation values for solutions
to the free $q$-de\-formed Schr\"{o}\-dinger equations [cf.
Eqs.~(\ref{FreParSch1N}) and (\ref{KonSchrGle}) in Chap.~\ref{LoeSchGleKap}].

We require that the solutions to the free $q$-de\-formed Schr\"{o}\-dinger
equations are subject to the following normalization condition:%
\begin{equation}
1=\frac{1}{2}\int\text{d}_{q}^{3}x\left(  \phi_{L}^{\ast}(\mathbf{x}%
,t)\circledast\phi_{R}(\mathbf{x},t)+\phi_{L}(\mathbf{x},t)\circledast\phi
_{R}^{\ast}(\mathbf{x},t)\right)  . \label{NorBed}%
\end{equation}
This condition is equivalent to%
\begin{equation}
1=\frac{1}{2}\int\text{d}_{q}^{3}\hspace{0.01in}p\left(  (c^{\ast
})_{\mathbf{p}}\circledast c_{\hspace{0.01in}\mathbf{p}}+c^{\hspace
{0.01in}\mathbf{p}}\circledast(c^{\ast})^{\mathbf{p}}\right)  .
\label{NorBedImp}%
\end{equation}
You can see this by inserting the expressions of
Eqs.~(\ref{EntWicEbeWelDreDim1}) and (\ref{EntWicEbeWelDreDim2}) into
Eq.~(\ref{NorBed}) and proceeding in the following manner:%
\begin{align}
&  \int\text{d}_{q}^{3}x\,\phi_{L}^{\ast}(\mathbf{x},t)\circledast\phi
_{R}(\mathbf{x},t)=\nonumber\\
&  \qquad=\int\text{d}_{q}^{3}p\int\text{d}_{q}^{3}p^{\prime}(c^{\ast
})_{\mathbf{p}}\circledast\hspace{-0.02in}\int\text{d}_{q}^{3}x\,(u^{\ast})_{\mathbf{p}%
}(\mathbf{x},t)\circledast u_{\hspace{0.01in}\mathbf{p}^{\prime}}%
(\mathbf{x},t)\circledast c_{\hspace{0.01in}\mathbf{p}^{\prime}}\nonumber\\
&  \qquad=\int\text{d}_{q}^{3}p\int\text{d}_{q}^{3}p^{\prime}(c^{\ast
})_{\mathbf{p}}\circledast\operatorname*{vol}\nolimits^{-1}\hspace
{-0.01in}\delta_{q}^{\hspace{0.01in}3}((\ominus\hspace{0.01in}\kappa^{-1}\mathbf{p)}%
\oplus\mathbf{p}^{\hspace{0.01in}\prime})\circledast c_{\hspace{0.01in}%
\mathbf{p}^{\prime}}\nonumber\\
&  \qquad=\int\text{d}_{q}^{3}p\,(c^{\ast})_{\mathbf{p}}\circledast
c_{\hspace{0.01in}\mathbf{p}}. \label{UmNorBedImp}%
\end{align}
Note that the last two steps of the above calculation follow from
Eqs.~(\ref{OrtRelEbeWel0Schr}) and (\ref{AlgChaIdeqDelFkt}) of
Chap.~\ref{KapOrtVolEBeWel}.

We continue with the expectation value of the momentum operator. We determine
this expectation value in position space as well as in momentum space:%
\begin{align}
\langle P^{A}\rangle_{\phi}  &  =\frac{1}{2}\int\text{d}_{q}^{3}x\left(
\phi_{L}^{\ast}(\mathbf{x},t)\circledast\text{i}^{-1}\partial^{A}%
\triangleright\phi_{R}(\mathbf{x},t)+\phi_{L}(\mathbf{x},t)\circledast
\text{i}^{-1}\partial^{A}\,\bar{\triangleright}\,\phi_{R}^{\ast}%
(\mathbf{x},t)\right) \nonumber\\
&  =\frac{1}{2}\int\text{d}_{q}^{3}p\left(  (c^{\ast})_{\mathbf{p}}\circledast
p^{A}\circledast c_{\hspace{0.01in}\mathbf{p}}+c^{\hspace{0.01in}\mathbf{p}%
}\circledast p^{A}\circledast(c^{\ast})^{\mathbf{p}}\right)  .
\label{ErwImpSchTeiFre1}%
\end{align}
To obtain the expression in momentum space from that in position space, we
proceed as follows:%
\begin{align}
&  \int\text{d}_{q}^{3}x\,\phi_{L}^{\ast}(\mathbf{x},t)\circledast
\text{i}^{-1}\partial^{A}\triangleright\phi_{R}(\mathbf{x},t)=\nonumber\\
&  \qquad=\int\text{d}_{q}^{3}p\int\text{d}_{q}^{3}p^{\prime}(c^{\ast
})_{\mathbf{p}}\circledast\int\text{d}_{q}^{3}x\,(u^{\ast})_{\mathbf{p}%
}(\mathbf{x},t)\circledast u_{\hspace{0.01in}\mathbf{p}^{\prime}}%
(\mathbf{x},t)\circledast p^{A}\circledast c_{\hspace{0.01in}\mathbf{p}%
^{\prime}}\nonumber\\
&  \qquad=\int\text{d}_{q}^{3}p\int\text{d}_{q}^{3}p^{\prime}(c^{\ast
})_{\mathbf{p}}\circledast\operatorname*{vol}\nolimits^{-1}\hspace
{-0.01in}\delta_{q}^{\hspace{0.01in}3}((\ominus\hspace{0.01in}\kappa^{-1}\mathbf{p)}%
\oplus\mathbf{p}^{\hspace{0.01in}\prime})\circledast p^{A}\circledast
c_{\hspace{0.01in}\mathbf{p}^{\prime}}\nonumber\\
&  \qquad=\int\text{d}_{q}^{3}p\,(c^{\ast})_{\mathbf{p}}\circledast
p^{A}\circledast c_{\hspace{0.01in}\mathbf{p}}. \label{UmErwImp}%
\end{align}
In the first step of the above calculation, we used the series expansion in terms of $q$-de\-formed plane
waves [see Eq.~(\ref{EntWicEbeWelDreDim1}) in the previous chapter]
and applied the eigenvalue equations for $q$-de\-formed plane waves [see
Eq.~(\ref{ImpEigWelSol}) in Chap.~\ref{LoeSchGleKap}]:%
\begin{align}
\text{i}^{-1}\partial^{A}\triangleright\phi_{R}(\mathbf{x},t)  &
=\int\text{d}_{q}^{3}p\,\text{i}^{-1}\partial^{A}\triangleright u_{\hspace
{0.01in}\mathbf{p}}(\mathbf{x},t)\circledast c_{\hspace{0.01in}\mathbf{p}%
}\nonumber\\
&  =\int\text{d}_{q}^{3}p\,u_{\hspace{0.01in}\mathbf{p}}(\mathbf{x}%
,t)\circledast p^{A}\circledast c_{\hspace{0.01in}\mathbf{p}}.
\end{align}
The further steps in Eq.~(\ref{UmErwImp}) correspond to those in
Eq.~(\ref{UmNorBedImp}).

The expectation value of the momentum operator behaves as follows under
conjugation:%
\begin{equation}
\overline{\langle P^{A}\rangle_{\phi}}=\langle P_{A}\rangle_{\phi}.
\end{equation}
This identity follows from the last expression in Eq.~(\ref{ErwImpSchTeiFre1})
if we take into account Eq.~(\ref{KonBedEntKoe}) of
Chap.~\ref{KapOrtVolEBeWel} as well as the conjugation properties of momentum
coordinates, $q$-integral, and star product [cf. Eq.~(\ref{KonEigSteProFkt})
in Chap.~\ref{KapQuaZeiEle} and Eq.~(\ref{KonEigVolInt}) in
Chap.~\ref{KapParDer}]:%
\begin{equation}
\overline{\int\text{d}_{q}^{3}p\,(c^{\ast})_{\mathbf{p}}\circledast
p^{A}\circledast c_{\hspace{0.01in}\mathbf{p}}}=\int\text{d}_{q}%
^{3}p\,c^{\hspace{0.01in}\mathbf{p}}\circledast p_{A}\circledast(c^{\ast
})^{\mathbf{p}}.
\end{equation}

We can also write down expressions for the expectation value of the position
operator. Concretely, we have%
\begin{align}
\langle X^{A}\rangle_{\phi}  &  =\frac{1}{2}\int\text{d}_{q}^{3}x\left(
\phi_{L}^{\ast}(\mathbf{x},t)\circledast x^{A}\circledast\phi_{R}%
(\mathbf{x},t)+\phi_{L}(\mathbf{x},t)\circledast x^{A}\circledast\phi
_{R}^{\ast}(\mathbf{x},t)\right) \nonumber\\
&  =\frac{1}{2}\int\text{d}_{q}^{3}p\left(  (c^{\ast})_{\mathbf{p}%
}(t)\circledast\text{i}\partial_{p}^{A}\,\bar{\triangleright}\,c_{\hspace
{0.01in}\mathbf{p}}(t)+c^{\hspace{0.01in}\mathbf{p}}(t)\circledast
\text{i}\partial_{p}^{A}\triangleright(c^{\ast})^{\mathbf{p}}(t)\right)
\label{ErwOrtSchTeiFre1}%
\end{align}
with%
\begin{align}
c_{\hspace{0.01in}\mathbf{p}}(t)  &  =\exp\left(  -\frac{\text{i}%
t\hspace{0.01in}\mathbf{p}^{2}}{2\hspace{0.01in}m}\right)  \circledast
c_{\hspace{0.01in}\mathbf{p}}, & (c^{\ast})_{\mathbf{p}}(t)  &  =(c^{\ast
})_{\mathbf{p}}\circledast\exp\left(  \frac{\text{i}t\hspace{0.01in}%
\mathbf{p}^{2}}{2\hspace{0.01in}m}\right)  ,\nonumber\\
c^{\hspace{0.01in}\mathbf{p}}(t)  &  =c^{\hspace{0.01in}\mathbf{p}}%
\circledast\exp\left(  \frac{\text{i}t\hspace{0.01in}\mathbf{p}^{2}}%
{2\hspace{0.01in}m}\right)  , & (c^{\ast})^{\mathbf{p}}(t)  &  =\exp\left(
-\frac{\text{i}t\hspace{0.01in}\mathbf{p}^{2}}{2\hspace{0.01in}m}\right)
\circledast(c^{\ast})^{\mathbf{p}}. \label{ZeiAbhDKoe}%
\end{align}
To derive the last expression in Eq.~(\ref{ErwOrtSchTeiFre1})\ from that in position space, we use the series expansion in terms of $q$-de\-formed plane
waves together with the eigenvalue equations%
\begin{align}
x^{A}\circledast\exp_{q}(\mathbf{x}|\text{i}\mathbf{p})  &  =\exp
_{q}(\mathbf{x}|\text{i}\mathbf{p})\,\bar{\triangleleft}\,\partial_{p}%
^{A}\text{i},\nonumber\\
\text{i}\partial_{p}^{A}\triangleright\exp_{q}(\text{i}^{-1}\mathbf{p}%
|\mathbf{x})  &  =\exp_{q}(\text{i}^{-1}\mathbf{p}|\mathbf{x})\circledast
x^{A},
\end{align}
and%
\begin{align}
\exp_{q}^{\ast}(\text{i}\mathbf{p}|\mathbf{x})\circledast x^{A}  &
=\text{i}\partial_{p}^{A}\,\bar{\triangleright}\,\exp_{q}^{\ast}%
(\text{i}\mathbf{p}|\mathbf{x}),\nonumber\\
\exp_{q}^{\ast}(\mathbf{x}|\text{i}^{-1}\mathbf{p})\triangleleft\partial
_{p}^{A}\text{i}  &  =x^{A}\circledast\exp_{q}^{\ast}(\mathbf{x}|\text{i}%
^{-1}\mathbf{p}).
\end{align}
So, we can proceed similarly as in Eq.~(\ref{UmErwImp}):%
\begin{align}
&  \int\text{d}_{q}^{3}x\,\phi_{L}^{\ast}(\mathbf{x},t)\circledast
x^{A}\circledast\phi_{R}(\mathbf{x},t)=\nonumber\\
&  \qquad=\int\text{d}_{q}^{3}p\int\text{d}_{q}^{3}p^{\prime}(c^{\ast
})_{\mathbf{p}}(t)\circledast\hspace{-0.02in}\int\text{d}_{q}^{3}x\,\text{i}\partial_{p}%
^{A}\,\bar{\triangleright}\,(u^{\ast})_{\mathbf{p}}(\mathbf{x},t)\circledast
u_{\hspace{0.01in}\mathbf{p}^{\prime}}(\mathbf{x},t)\circledast c_{\hspace
{0.01in}\mathbf{p}^{\prime}}(t)\nonumber\\
&  \qquad=\int\text{d}_{q}^{3}p\,(c^{\ast})_{\mathbf{p}}(t)\circledast
\text{i}\partial_{p}^{A}\,\bar{\triangleright}\,\operatorname*{vol}%
\nolimits^{-1}\hspace{-0.02in}\int\text{d}_{q}^{3}p^{\prime}\delta_{q}%
^{\hspace{0.01in}3}((\ominus\hspace{0.01in}\kappa^{-1}\mathbf{p)}\oplus\mathbf{p}%
^{\hspace{0.01in}\prime})\circledast c_{\hspace{0.01in}\mathbf{p}^{\prime}%
}(t)\nonumber\\
&  \qquad=\int\text{d}_{q}^{3}p\,(c^{\ast})_{\mathbf{p}}(t)\circledast
\text{i}\partial_{p}^{A}\,\bar{\triangleright}\,c_{\hspace{0.01in}\mathbf{p}%
}(t).
\end{align}

For the sake of completeness, we note that the expectation value $\langle
X^{A}\rangle_{\phi}$ behaves under conjugation as follows:%
\begin{equation}
\overline{\langle X^{A}\rangle_{\phi}}=\langle X_{A}\rangle_{\phi}.
\end{equation}
This identity follows from the expression for $\langle X^{A}\rangle_{\phi}$ in
position space if we take into account Eq.~(\ref{VerKonWelFkt}) in
Chap.~\ref{LoeSchGleKap} as well as the conjugation properties of spatial
coordinates, $q$-integral, and star product.

Also note that the expectation value $\langle X^{A}\rangle_{\phi}$ is usually
not time-independent, unlike the expectation value $\langle P^{A}\rangle
_{\phi}$. You can see this from the expressions for $\langle X^{A}%
\rangle_{\phi}$ and $\langle P^{A}\rangle_{\phi}$ in momentum space [cf.
Eqs.~(\ref{ErwImpSchTeiFre1}) and (\ref{ErwOrtSchTeiFre1})].

{\normalsize
\bibliographystyle{abbrv}
\bibliography{book,habil}
}

\end{document}